\documentclass[12pt,english]{article}
\pdfoutput=1
\usepackage[T1]{fontenc}
\usepackage{amssymb}
\usepackage{amsmath}
\usepackage{dsfont}
\usepackage{bbm}
\usepackage{cite}
\usepackage{epsfig}
\usepackage{scalerel}
\usepackage[unicode=true,bookmarks=true,bookmarksnumbered=false,bookmarksopen=false,
 breaklinks=false,pdfborder={0 0 1},backref=false,colorlinks=true]
 {hyperref}
\usepackage[hang,flushmargin]{footmisc}
\usepackage{color}
\usepackage{shuffle}
\usepackage{cleveref}
\usepackage{nicematrix}
\usepackage{braket}
\usepackage{booktabs}

\usepackage[latin1]{inputenc}
\usepackage{amsmath}
\usepackage{amsfonts}
\usepackage{amssymb}

\usepackage{graphicx}
\usepackage[export]{adjustbox}

\usepackage{tikz}
\usetikzlibrary{arrows.meta}
\usetikzlibrary{decorations.markings}
\tikzset{
        internalleg/.style={postaction={decorate},
		very thick,
        decoration={markings, mark=at position 0.5 with {\arrow{stealth}}}
    },
        leafleg/.style={postaction={decorate},
		very thick,
        Ellipse-,
        decoration={markings, mark=at position 0.5 with {\arrow{stealth}}}
    },
        leaflegs/.style={postaction={decorate},
		very thick,
        Ellipse-,
        decoration={markings, mark=at position 0.8 with {\arrow{stealth}}}
    },
        rootleg/.style={postaction={decorate},
		very thick,
        -to reversed,
        decoration={markings, mark=at position 0.5 with {\arrow{stealth}}}
    }
}

\usepackage[most]{tcolorbox}

\usepackage[bbgreekl]{mathbbol}
\DeclareMathSymbol\bbDelta  \mathord{bbold}{"01}

\makeatletter
\def\simgt{\mathrel{\lower2.5pt\vbox{\lineskip=0pt\baselineskip=0pt
           \hbox{$>$}\hbox{$\sim$}}}}
\def\simlt{\mathrel{\lower2.5pt\vbox{\lineskip=0pt\baselineskip=0pt
           \hbox{$<$}\hbox{$\sim$}}}}
\makeatother

\newcommand{\be}{\begin{equation}}
\newcommand{\ee}{\end{equation}}
\newcommand{\bea}{\begin{eqnarray}}
\newcommand{\eea}{\end{eqnarray}}

\usepackage{xparse}

\Crefname{equation}{Eq.}{Eqs.}

\ifdefined \Ref
\renewcommand{\Ref}[1]{Ref.~\cite{#1}}
\else
\newcommand{\Ref}[1]{Ref.~\cite{#1}}
\fi

\newcommand{\Eq}[1]{Eq.~\eqref{#1}}

\newcommand{\Sec}[1]{Sec.~\ref{#1}}

\newcommand{\App}[1]{App.~\ref{#1}}

\newcommand{\tr}{\textrm{tr}}

\newcommand{\ov}{\overline}

\newcommand{\qandq}{\qquad \textrm{and} \qquad}

\newcommand{\D}{\mathbbm{D}}

\renewcommand{\O}{{\cal O}}
\renewcommand{\L}{{\cal L}}
\newcommand{\K}{K}

\newcommand{\M}{{\cal M}}

\newcommand{\Cont}{\Delta}
\newcommand{\Amp}{{A}}
\newcommand{\AmpD}{\mathbbm{A}}

\newcommand{\fddu}[3]{f_{ #1#2}^{\phantom{#1 #2} #3}}

\newcommand{\fbarddu}[3]{f_{ \ov #1 \ov #2}^{\phantom{\ov#1 \ov #2} \ov #3}}

\newcommand{\fudd}[3]{f^{#1}_{\, \phantom{#1 } #2 #3}}

\newcommand{\Fddu}[3]{F_{#1#2}^{\phantom{#1 #2} #3}}

\newcommand{\Fuud}[3]{F^{#1#2}_{\phantom{#1 #2} #3}}
\newcommand{\Fudd}[3]{F^{#1}_{\phantom{#1 } #2 #3}}

\newcommand{\lr}[2]{\overset{\leftrightarrow}{#1}_{#2}}
\newcommand{\lrsq}[1]{\overset{\leftrightarrow}{#1}^{\lower.5em \hbox{\tiny $2$}}}

\hypersetup{citecolor=blue,linkcolor=blue,urlcolor=blue}

\numberwithin{equation}{section}

\begin{document}

\DeclareDocumentCommand\eq{mg}{\be\begin{aligned}#1\IfNoValueF{#2}{\label{#2}}\end{aligned}\ee}

\interfootnotelinepenalty=10000
\baselineskip=18pt

\hfill   CALT-TH-2022-002

\vspace{2.5cm}
\thispagestyle{empty}

\begin{center}

{\Large \bf
On-shell Correlators and Color-Kinematics \\ \medskip    Duality in Curved Symmetric Spacetimes \\
}
\bigskip\vspace{1cm}{
{\large Clifford Cheung${}^{1}$, Julio Parra-Martinez${}^{1}$, and Allic Sivaramakrishnan${}^{2}$}
} \\[7mm]
{\it ${}^{1}$Walter Burke Institute for Theoretical Physics\\[-1mm]
    California Institute of Technology, Pasadena, CA 91125  \\
    ${}^{2}$Department of Physics and Astronomy \\[-1mm]
		University of Kentucky, Lexington, KY 40506 }
 \let\thefootnote\relax\footnote{\noindent 
\url{clifford.cheung@caltech.edu},
\url{jparram@caltech.edu},
\url{allicsiva@uky.edu}
} \\
 \end{center}

\bigskip
\centerline{\large\bf Abstract}
\begin{quote} \small

We define a perturbatively calculable quantity---the on-shell correlator---which furnishes a unified description of particle dynamics in curved spacetime.  Specializing to the case of flat and anti-de Sitter space, on-shell correlators coincide precisely with on-shell scattering amplitudes and boundary correlators, respectively. Remarkably, we find that symmetric manifolds admit a generalization of on-shell kinematics in which the corresponding momenta are literally the isometry generators of the spacetime acting on the external kinematic data.  These isometric momenta are intrinsically non-commutative but exhibit on-shell conditions that are identical to those of flat space, thus providing a common language for computing and representing on-shell correlators which is agnostic about the underlying geometry.
Afterwards, we compute tree-level on-shell correlators for biadjoint scalar (BAS) theory and the nonlinear sigma model (NLSM) and learn that color-kinematics duality is manifested at the level of fields under a mapping of the color algebra to the algebra of gauged isometries on the spacetime manifold.  Last but not least, we present a field theoretic derivation of the fundamental BCJ relations for on-shell correlators following from the existence of certain conserved currents in BAS theory and the NLSM.

\end{quote}

\setcounter{footnote}{0}

\setcounter{tocdepth}{2}
\newpage
\tableofcontents    
\newpage

\section{Introduction}

The past several decades have witnessed a resurgence of interest in the theory of scattering.  This work has galvanized new discoveries relevant to collider physics \cite{Bern:1994zx,Bern:1994cg,Bern:1995db,Bern:1997sc,Britto:2004nc,Berger:2008sj,Badger:2017jhb,Abreu:2017hqn}, the underlying structure of gauge theories \cite{Alday:2007hr,Bern:2007ct,Bern:2012di,Arkani-Hamed:2012zlh,Arkani-Hamed:2013jha,Carrasco:2021otn}, the ultraviolet properties of supergravity \cite{Bern:2012uf,Bern:2012cd,Bern:2012gh,Bern:2013qca,Bern:2013uka,Bern:2014sna,Bern:2014lha,Bern:2017lpv,Bern:2017ucb,Bern:2018jmv,Herrmann:2018dja,Bourjaily:2018omh,Edison:2019ovj}, and more recently, the black hole binary inspiral problem \cite{Cheung:2018wkq,Kosower:2018adc,Bern:2019nnu,Antonelli:2019ytb,Bern:2019crd,Bern:2021dqo,Herrmann:2021lqe,Herrmann:2021tct,Bern:2021yeh}.  At the same time, the study of scattering amplitudes has revealed hidden features endemic to a broad class of quantum field theories, including color-kinematics duality \cite{Bern:2008qj} and the double copy \cite{Kawai:1985xq,Bern:2010ue} (see \cite{BCJReview} for a detailed review).

Notably, the vast majority of these developments have relied critically on the acute simplifications which arise from on-shell kinematics in flat space. This is by no means an accident---off-shell quantities are oftentimes unphysical since they are, by and large, not invariant under changes of gauge or field basis.   As a result, one might expect that the many remarkable structures observed in flat space scattering have no vestiges in curved spacetime, where there is no obvious generalization of on-shell kinematics.  

Of course, many advances in this direction have been made for maximally symmetric spaces, namely anti-de Sitter (AdS) space and de Sitter (dS) space. Ongoing explorations of Mellin space \cite{Penedones10, FitzpatrickKPRV11,RastelliZ16,RastelliZ17,Sleight19,SleightT19,SleightT20, SleightT21}, transition amplitudes \cite{Giddings99,BalasubramanianGL99,Raju10}, harmonic analysis \cite{CostaGP14, LiuPRS18, GiombiST17,DiPietro21,HogervorstPV21}, and unitarity \cite{FitzpatrickK11,AldayC17,MeltzerPS19,Ponomarev19,MeltzerS20,Meltzer21,Meltzer:2021bmb,BaumannCDJLP21,Goodhew:2020hob,Jazayeri:2021fvk,Melville:2021lst,Goodhew:2021oqg} have all identified various notions of simplified and on-shell  kinematics for curved spacetime. Despite substantial progress, however, sophisticated techniques are still required in all but the simplest settings even to obtain tree-level integrands and study their properties.

In this paper, we generalize familiar concepts from flat space such as momentum, on-shell kinematics, and on-shell scattering amplitudes to a broad class of curved geometries.  Our analysis hinges on the fact that for a generic symmetric manifold---which is any spacetime whose curvature tensor is covariantly constant---one can elegantly reframe geometry and kinematics in terms of the generators of the underlying isometries of the spacetime. In this formulation we eschew the usual coordinate basis of the tangent space of the manifold, $\partial_\mu$, instead opting for an alternative frame in which directions are labeled by an overcomplete basis of Killing vectors, $K_A = K_A^\mu\partial_\mu$.   We dub all quantities expressed in terms of these generators as ``isometric''.

Armed with an isometric frame for the geometry, we derive a new formulation of isometric kinematics applicable to curved spacetime.   In this picture, flat space momentum $p_\mu$ is generalized to an isometric momentum $\D_A$ that is literally the isometry-generating Lie derivative corresponding to the Killing vector $K_A$.  Isometric momenta are inherently non-commutative, and so the commutator $[\D_A, \D_B] = \Fddu{A}{B}{C} \D_C$ encodes the Killing algebra.  Meanwhile, any isometry invariant function automatically manifests a generalization of total momentum conservation.  By defining a set of zero eigenfunctions of the Killing Casimir, $\D^2 = \D_A \D^A$, we also obtain a generalization of on-shell wavefunctions for external massless particles.  

Altogether, this approach furnishes a curved spacetime formulation of on-shell kinematics which is literally identical to flat space---complete with isometric momentum conservation and isometric on-shell conditions---except that it is non-commutative.  Armed with this insight, we  propose a generalized observable that is calculable in curved geometries, dubbed the ``on-shell correlator''.  At tree-level, the on-shell correlator is simply a solution to the classical equations of motion that asymptotes to a superposition of on-shell wavefunctions.  Equivalently, the on-shell correlator is a standard position space correlator with all external propagators amputated and then dressed with on-shell wavefunctions.  Specializing to flat space and AdS, on-shell correlators reduce, by construction, to on-shell scattering amplitudes and boundary correlators, respectively.  While our analysis will focus on on-shell correlators for symmetric spacetimes, these objects can be defined more generally.

\begin{table}[t]
        \centering
        \begin{tabular}{|c|c|c|}
        \midrule
         flat  space                   & AdS                     &  symmetric space        \\\midrule
        \quad on-shell scattering amplitude \quad & boundary correlator      & on-shell correlator            \\
         plane wave                    & \quad bulk-boundary propagator \quad & wavefunctions                  \\
         translations                  & conformal group          & isometries                       \\
         momentum                      & conformal generator      &\quad  isometric momentum  \quad           \\ \midrule
        \end{tabular}
        \caption{Dictionary for relevant quantities in flat space, AdS, and general symmetric spaces. }
\label{tab:fAdSdS}
\end{table}

As we will show, in broad class of theories, every on-shell correlator can be written as a differential operator acting on an integrated product of wavefunctions that corresponds to a momentum-conserving delta function in flat space and to a boundary correlator for a contact interaction in AdS.   The differential operator is a function of isometry generators acting on the  kinematic data labeling the external legs of the on-shell correlator.  This differential operator is identical to what would be computed via flat space Feynman diagrams, except maintaining the relative ordering of momenta.   Propagators are given by the inverse Killing Casimir, $1/\D^2$, while vertices are functions of the isometric momenta $\D_A$. The specialization of some of the relevant concepts to flat space and AdS is summarized in Tab.~\ref{tab:fAdSdS}.  These results are essentially a generalization of \cite{DiwakarHRT21, EberhardtKM20}, albeit lifted from AdS embedding space to a much larger class of symmetric spacetimes.

Conveniently, the on-shell correlators computed in local theories automatically have propagator denominators that fully commute with each other as well as with their numerators. Thus, locality enforces a well-defined separation between numerator and denominator factors.   For this reason it is natural to ask whether structures like color-kinematics duality and the double copy---which hinge critically on this split---can be generalized to curved spacetime.  As we will also prove, this is indeed the case.  

In particular, we build on the results of \cite{CCK}, which presented a  quantum field theoretic derivation of the color-kinematics duality that maps biadjoint scalar (BAS) theory to the nonlinear sigma model (NLSM).  In flat space, this duality sends the color algebra to the diffeomorphism algebra at the level of fields in the equations of motion.  In the present work, we essentially repeat this analysis for BAS and the NLSM in isometric frame.  By inspection, we find that color-kinematics duality corresponds to substituting the algebra of color for the algebra of gauged isometries of the underlying manifold.

Last but not least, we show how conservation of the kinematic current \cite{CCK}--- which directly mandates the Jacobi identity---directly implies the fundamental BCJ relation for all $n$-point on-shell correlators of BAS theory and NLSM in curved spacetime.  This furnishes a field theoretic proof of the conjecture of \cite{DiwakarHRT21} for the special case of AdS.  For other approaches to color-kinematics duality in curved spacetime at four points see \cite{BzowskiMS17,FarrowLM18,LipsteinM19,Zhou21,AldayBFZ21,AlbayrakKM20,Sivaramakrishnan21,ArmstrongLM20}, and at five points see \cite{Alday:2022lkk}.

\medskip

\noindent {\bf Note added:}  During the completion of this work we learned of related ongoing work by Herderschee, Roiban, and Teng in \cite{HRTpreprint}, which focuses on the case of AdS but overlaps with our discussion of the isometric representation (which they dub the differential representation in the original reference \cite{DiwakarHRT21}) and Berends-Giele recursion. We are grateful to those authors for useful discussions and sharing their preprint with us.

\section{Geometry}
\label{sec:geo}

In this section we briefly review the geometric description of isometries in curved spacetime.  We then show that for (irreducible) symmetric spacetimes, all geometric quantities can be mechanically transformed to an  isometric frame in which coordinates are labeled by the underlying isometry generators of the manifold itself.  In this frame, tensors carry  isometric indices that run over isometry generators rather than directions in spacetime.  Armed with this notion, we then describe how to recast a broad class of scalar field theory Lagrangians into isometric frame.

\subsection{Preliminaries}

To begin, let us review some basic facts about the geometry of a manifold $\M$.  The isometries of ${\cal M}$ are generated by the set of Killing vectors,
\eq{
\K_A = K_A^\mu \partial_\mu,
}  
whose commutator yields the Killing algebra
\eq{
{} [K_A, K_B] = \Fddu{A}{B}{C} K_C,
}{Killing_algebra}
where $\Fddu{A}{B}{C}$ is the structure constant.  In components, the Killing algebra is
\eq{
\K_A^\mu \nabla_\mu \K_B^\nu - \K_B^\mu \nabla_\mu \K_A^\nu = \Fddu{A}{B}{C} \K_C^\nu.
}
Note that covariant and partial derivatives are interchangeable in these commutators since the difference involves terms proportional to the connection that always cancel identically.   Let us also define the Killing metric, 
\eq{
g_{AB} =  \Fddu{A}{C}{D}\Fddu{B}{D}{C} ,
}{def_gAB}
which is the natural bilinear form in this space, and which is used to freely raise and lower indices.   Note that the structure constant with all lowered indices, $F_{ABC}$, is fully antisymmetric.\footnote{The structure constant $\Fddu{A}{B}{C}$ is by definition antisymmetric in its first and second index.  Meanwhile, since $F_{ABC}=
\Fddu{A}{B}{D} \Fddu{D}{E}{F}\Fddu{C}{F}{E} = \Fddu{B}{E}{D} \Fddu{A}{D}{F}\Fddu{C}{F}{E}-\Fddu{A}{E}{D} \Fddu{B}{D}{F}\Fddu{C}{F}{E}$, we see that $F_{ABC}=F_{BCA}=F_{CAB}$ is also cyclically symmetric, thus implying full antisymmetry.  }

For our analyses it will be convenient to also define a shorthand for the isometry-generating Lie derivative with respect to $K_A$,
\eq{
\D_A = {\cal L}_{\K_A}.
}{def_DA}
Recall that the Lie derivative acts on an arbitrary tensor as
\eq{
\D_{A} \O^{\rho_1 \rho_2 \rho_3 \cdots \rho_n}_{\phantom{\rho_1 \rho_2 \rho_3 \cdots \rho_n} \sigma_1 \sigma_2 \cdots \sigma_m} = & \phantom{{}-{}} \K^\mu_A \nabla_\mu \O^{\rho_1 \rho_2 \rho_3 \cdots \rho_n}_{\phantom{\rho_1 \rho_2 \rho_3 \cdots \rho_n} \sigma_1 \sigma_2\sigma_3 \cdots \sigma_m} \\
&- \nabla_\mu \K^{\rho_1}_A  \O^{\mu \rho_2 \rho_3 \cdots \rho_n}_{\phantom{\mu \rho_2 \rho_3 \cdots \rho_n} \sigma_1 \sigma_2\sigma_3 \cdots \sigma_m} - \nabla_\mu \K^{\rho_2}_A  \O^{\rho_1 \mu \rho_3 \cdots \rho_n}_{\phantom{\rho_1 \mu \rho_3 \cdots \rho_n} \sigma_1 \sigma_2 \sigma_3 \cdots \sigma_m} - \cdots \\
&+ \nabla_{\sigma_1} \K^{\mu}_A  \O^{\rho_1 \rho_2 \rho_3 \cdots \rho_n}_{\phantom{\rho_1 \rho_2 \rho_3 \cdots \rho_n} \mu \sigma_2 \sigma_3 \cdots \sigma_m} + \nabla_{\sigma_2} 
\K^{\mu}_A  \O^{\rho_1 \rho_2  \rho_3\cdots \rho_n}_{\phantom{\rho_1 \rho_2 \rho_3 \cdots \rho_n} \sigma_1 \mu \sigma_3 \cdots \sigma_m} + \cdots ,
 }{DO_tensor}
 where covariant and partial derivatives are again interchangeable.
By definition, the metric $g_{\mu\nu}$ on ${\cal M}$ is annihilated by the Lie derivative with respect to a Killing vector $K_A$, 
\eq{
0 = \D^A g_{\mu\nu}= \nabla_{(\mu} \K_{\nu)}^A ,
}{Killing_equation}
which is the Killing equation.
Contracting this with the metric, we obtain
\eq{
\nabla^\mu K_\mu^A=0,
}{dK_zero}
so the Killing vectors have vanishing divergence.  Note that the Killing algebra is also encoded in  various commutators of the Lie derivative and the Killing vector, so for example
\eq{
{} [\D_A, \D_B] = \Fddu{A}{B}{C} \D_C \qandq [\D_A , K_B^\mu]  =\D_A K_B^\mu  = \Fddu{A}{B}{C} K_C^\mu,
}{Killing_algebra_extra}
which will come in handy in technical calculations later on.

 By contracting a pair of Lie derivatives with the Killing metric, we obtain the quadratic Casimir of the Killing algebra,
\eq{
\D^2 =  \D_A \D^A = g^{AB} \D_A \D_B ,
}
which trivially commutes with all Lie derivatives 
\eq{
{} [\D_A, \D^2 ] &= 0, 
}
and is thus an invariant.  Later on, we will see that states are naturally classified by their eigenvalue under the Killing Casimir.

\subsection{Isometric Frame}

Hereafter, let us assume that the spacetime manifold $\M$ is symmetric, which means it is a coset space $G/H$ where $G$ is a Lie group and $H$ is a compact subgroup of $G$. Symmetric spacetimes have constant curvature and exhibit a set of Killing vectors, $K_A$, with $A=1,\ldots, {\rm dim}(G)$, which descend from the right-invariant vector field on the group manifold $G$. At any given point $x \in \M$,  precisely ${\rm dim}(H)$ of these Killing vectors are zero, while the remaining ${\rm dim }(G/H)$ span the tangent space of $\M$. Throughout, we will make the additional assumption that the generators of $G/H$ furnish an irreducible representation of $H$.

In the more familiar language of spontaneous symmetry breaking, the Killing vectors in $H$ and $G/H$ correspond to unbroken and broken generators, respectively.  By definition, the ${\rm dim}(H)$ broken generators leave the point $x\in \M$ invariant, and thus the corresponding Killing vectors vanish at $x$, while the ${\rm dim}(G/H)$ broken generators act nontrivially.
If we denote the generators of $H$ by $\cal T$ and the elements spanning $G/H$ by $\cal X$, the commutator algebra of the corresponding Killing vectors has the schematic structure
\eq{
        {}   [ {\cal T}, {\cal T}] \sim {\cal T}\,, \qquad [{\cal T}, {\cal X}] \sim {\cal X}\,, \qquad [{\cal X}, {\cal X}] \sim {\cal T}\,,
}
which has the defining automorphism of a symmetric space 
\eq{
{\cal T} \to {\cal T},\qquad {\cal X} \to - {\cal X}\,.
}{eq:symz2}
A crucial byproduct of this fact is that the structure constants of the Killing algebra, $F_{ABC}$, only have nonzero components when the indices $A,B,C,$ correspond either to three unbroken generators or to one unbroken and two broken generators.   All other combinations of generators are forbidden by the underlying parity of the symmetric manifold.  As a consequence, we learn that
 the structure constants automatically vanish when fully contracted into Killing vectors,
\eq{
  F^{ABC}  \K_A^\mu \K_B^\nu K_C^\rho = 0.
}{FKKK}
This happens because the parity in \Eq{eq:symz2} enforces that at least one of the Killing vectors in \Eq{FKKK} is an unbroken generator, which must in turns be zero because it acts trivially on the point $x\in \M$.
 A similar identity holds if we replace a Killing vector with a Lie derivative acting on a scalar,
\eq{
  F^{ABC}  \K_A^\mu \K_B^\nu \D_C = 0,
}{FKKD}
since in this case $\D_A = K_A^\mu \nabla_\mu$ and we can use \Eq{FKKK}.

Given our stated assumptions, it is possible to express the metric directly in terms of the Killing vectors via \cite{Boulware:1981ns,Camporesi:1990wm}
\eq{
g_{\mu\nu} = g_{AB}  \K_{\mu}^A \K_{\nu}^B ,
}{g_from_K}
which automatically satisfies the Killing equation in \Eq{Killing_equation}.  
It is obvious that \Eq{g_from_K} is only possible if the number of Killing vectors is equal or greater than the dimensionality of the manifold, which is indeed the case when $\M$ is symmetric since 
\eq{
   {\rm dim}(G) > {\rm dim}(\M) = {\rm dim }(G/H) .
}{rank_inequality}
Notably, \Eq{g_from_K} is closely related to the definition of a local frame, {\it i.e.}, a tetrad, but with the critical difference that the tetrad basis has been replaced with an overcomplete set of Killing vectors. In spite of this, it is should be clear that \Eq{g_from_K} can be   mechanically inserted into expressions in order to transform canonical metric indices labeled by $\mu, \nu,\rho$, etc., into isometric indices labeled by $A, B, C$, etc.  We refer to this procedure as a transformation to isometric frame, precisely in analogy with the more conventional transformation to tetrad frame.  
In isometric frame, the indices of all tensors are of isometric type and run over the basis of isometries exhibited by the spacetime.

We emphasize that our setup is different from the usual construction of a local frame in a symmetric space \cite{Camporesi:1990wm}, which is given by regarding $G$ as an $H$-bundle over $G/H$ and pulling back the right-invariant Killing vectors on $G$ using an arbitrary section. In the usual construction, the section picks out the ${\rm dim }(G/H)$ non-vanishing Killing vectors at each point in $\M$, whereas our construction employs a global overcomplete basis.  Said another way, \Eq{g_from_K} does not depict a bona fide change of frame, since
the inverse relation of \Eq{g_from_K},
\eq{
 g_{AB} \neq g_{\mu\nu} \K^\mu_A \K^\nu_B,
 }
fails on account of \Eq{rank_inequality}, which stipulates that the rank of $g_{AB}$ exceeds the dimensionality of the basis of Killing vectors $K_A$.   

It is illuminating to consider the mapping to isometric frame in a simple case, for example when $\M$ is the $n$-sphere.  Given any point $x\in \M$, the group $G$ is the group of $(n+1)$-dimensional rotations while $H$ is generated the group of $n$-dimensional rotations that leave $x$ invariant.  At the point $x\in \M$, the unbroken generators of $H$ are identically zero while the remaining broken generators of $G/H$ are non-trivial.  The corresponding Killing Casimir is simply the rotation Casimir.
A second example is when $\M$ is flat space.  For any point $x\in \M$, the group $G$ is the Poincare group, while the subgroup $H$ is the Lorentz group defined about the origin $x$, which leaves this point invariant.  Hence, the broken generators of $G/H$ are translations, as expected.  In this case the Killing Casimir is that of the Poincare group, which is the d'Alembertian.   Another broad class of examples is the set of maximally symmetric spaces, which can be elegantly recast in terms of the embedding space formalism.  See \App{sec:embed} for details.

Last but not least, let us demonstrate mechanically how to transform a field theory on a symmetric spacetime manifold to isometric frame. 
As we will show, this procedure is always possible in an arbitrary theory of scalars with at most one derivative per field.   For the present work, we will focus entirely on this broad class of theories.  
The Lagrangian for this theory is
\eq{
\L(\phi, \nabla_\mu \phi) = \frac{1}{2} \nabla_\mu \phi \nabla^\mu \phi + \cdots.
}{L_gen}
The corresponding Euler-Lagrange equations of motion are second order in derivatives,
\eq{
0=\nabla_\mu \left(\frac{\delta \L}{\delta \nabla_\mu \phi}\right) - \frac{\delta \L}{\delta \phi} = \nabla^2 \phi + \cdots,
}{EOM_gen} 
so the initial value problem is well-posed and ghost modes are absent.
Using \Eq{g_from_K}, it is straightforward to recast all covariant derivatives in terms of Lie derivatives,
\eq{
\L(\phi, \D_A\phi) = \frac{1}{2}\D_A \phi \D^A \phi + \cdots .
}{L_gen_D}
Since the Lie derivative satisfies the product rule just like the covariant derivative, the derivation of the Euler-Lagrange equations of motion for \Eq{L_gen_D} is unchanged\footnote{Here we have used that a total Lie derivative of a scalar can be rewritten as a total partial derivative.} and we obtain
\eq{
0=\D_A \left(\frac{\delta \L}{\delta \D_A \phi}\right) - \frac{\delta \L}{\delta \phi} = \D^2 \phi + \cdots.
}{EOM_gen_D}
Comparing the equations of motion in \Eq{EOM_gen} and \Eq{EOM_gen_D} we see that the Laplacian is equivalent to the Lie derivative squared, {\it i.e.}~the Killing Casimir, when acting on a scalar.  This equivalence can also be derived more directly,
\eq{
\nabla^2 \phi &= \nabla_\mu( g^{\mu\nu} \nabla_\nu \phi) =  g^{AB} \nabla_\mu( K^{\mu}_A K^{\nu}_B \nabla_\nu \phi )
= g^{AB}  K^{\mu}_A \nabla_\mu(  K^{\nu}_B \nabla_\nu \phi ) = \D^2 \phi,
}{Laplacian_Lie}
where we have used the relationship between the metric and the Killing vectors in \Eq{g_from_K}.

The equivalence of the Laplacian and the Killing Casimir implies a simple but deep fact about perturbation theory on symmetric manifolds: the usual  propagator in curved spacetime, $1/\nabla^2$, is formally equivalent to the inverse Killing Casimir $1/\D^2$.  Though still a differential operator, the latter is maximally commutative.  As a result, this observation will enable a substantial simplification of perturbative calculations.

The above analysis relied on the fact that we assumed a scalar theory with at most one derivative per field.  When there is more than one derivative per field it is not obvious whether the corresponding Lagrangian can be transformed to isometric frame, {\it i.e.}~whether all covariant derivatives can be recast purely in terms of isometry-generating Lie derivatives.  We leave this question to future work.

\section{Kinematics}

In this section we generalize on-shell kinematics from flat space to curved symmetric manifolds.  By transforming to isometric frame, we will discover that powerful notions from flat space such as momentum conservation and on-shell conditions all have isometric cousins.

\subsection{Wavefunctions}

To begin, let us define on-shell wavefunctions to be sourceless solutions to the linearized scalar equation of motion.  In the present work we focus only on massless particles, but our results trivially generalize in the obvious way to the massive case. For a massless scalar particle labeled by $i$, the corresponding wavefunction $\psi(p_i,x)$ satisfies
\eq{
\nabla^2 \psi(p_i,x) = \D^2 \psi(p_i,x) =0,
}{wavefunction_def}
where $p_i$ is a kinematic label for this particular solution.  We will interpret $p_i$ as parameterizing the kinematic quantum numbers of the external state.    

Note that massless solutions need not exist in general, for instance if the geometry mandates a gap.  However, it is natural to expect massless solutions in the case that the spacetime manifold has non-compact directions.  In addition, even though \Eq{wavefunction_def} is the defining equation for a massless scalar, the generalization to higher spin is more subtle, as described in the special case of AdS in \cite{DiwakarHRT21}.  We leave an analysis of this higher spin question to future work.

As summarized in Tab.~\ref{tab:fAdSdS}, in flat space the Lie derivative $\D \sim \partial$ is a generator of spacetime translations and $\D^2 = \Box$ is the translation Casimir, {\it i.e.}~the d'Alembertian.  The wavefunction is a massless plane wave, $\psi(p_i,x) \sim e^{ip_ix}$ where $p_i^2=0$ is an on-shell momentum vector.  In AdS, the Lie derivative $\D$ is a conformal generator and $\D^2$ is the conformal Casimir.  The wavefunction is the bulk-boundary propagator for a massless scalar, $\psi(p_i,x) \sim K(p_i,x)$, where $p_i$ labels the point on the boundary.

The spacetime manifold is assumed to be symmetric, so it exhibits a set of isometries.  As a consequence, any solution to the equations of motion will map to other solutions under these isometries.  Mathematically, this implies that the Lie derivative $\D$ acting on the coordinate $x$ in the wavefunction can always be compensated by a coincident infinitesimal isometry transformation $\D_i$ acting on $p_i$, so
\eq{
(\D+ \D_{i}) \psi(p_i,x) = 0.
}{D_annihilates_psi}
We will refer to $\D_i$ as the isometric momentum of leg $i$, and it should be manipulated like a differential operator.  Note that we can be agnostic about the precise representation of $p_i$ under the isometry group.  All that matters is that commutators of isometric momenta exhibit the Killing algebra, so
\eq{
{} [ \D_{iA} ,\D_{iB} ] = \Fddu{A}{B}{C} \D_{iC} .
}
\Eq{wavefunction_def} and \Eq{D_annihilates_psi} together imply that
\eq{
\D_i^2 \psi(p_i,x)=0,
}{onshell_def}
which is the analog of the on-shell condition in flat space and the vanishing of the conformal Casimir in AdS.

\subsection{On-Shell Constraints}

We are now equipped to study the kinematics of multiple particles in the language of isometries.  First,  note that the product rule trivially implies an analog of momentum conservation.  To illustrate this, let us define a product wavefunction
\eq{
\psi(p_1, \cdots, p_n,x) = \prod_{i=1}^n \psi(p_i, x),
}
which is annihilated by the sum of derivatives,
\eq{
\left(\D+ \sum_{i=1}^n \D_{i} 
\right)\psi(p_1, \cdots, p_n,x) =0.
}{psi_annihilate}
When a function such as the product wavefunction is annihilated as in \Eq{psi_annihilate}, we will describe it as isometric invariant.  Furthermore, we refer to \Eq{psi_annihilate} and similar conditions as isometric momentum conservation, for obvious reasons.   The physical interpretation of \Eq{psi_annihilate} is that the momentum associated with the isometry flows through the coordinate $x$ and out through $p_i$.  
This fact motivates us to consider the spacetime integral of the product wavefunction, 
\eq{
\Cont_n = \Cont(p_1, \cdots, p_n) = \int_x \psi(p_1, \cdots, p_n,x),
}
which is a generalization of the $n$-point momentum-conserving delta function in flat space.  We dub this quantity the $n$-point contact correlator, since it coincides with the Witten diagram for a contact interaction in AdS. 

As expected, the contact correlator is annihilated by
\eq{
\left( \sum_{i=1}^n \D_{i} \right) \Cont_n=0,
}{cond_momentum}
so it is also isometric invariant and satisfies isometric momentum conservation.
Note also that \Eq{onshell_def} implies that
\eq{
\D_i^2 \Cont_n=0,
}{cond_onshell}
which we dub the isometric on-shell condition.  
Obviously, \Eq{cond_momentum} and \Eq{cond_onshell} are the generalizations of total momentum conservation and on-shell conditions.  Consequently, when acting on the contact correlator, the Lie derivatives are effectively curved space momentum variables which behave almost exactly the same as on-shell momenta in flat space.   

The only substantive difference which arises in curved spacetime is that the isometric momenta are non-commutative, exhibiting the Killing algebra
\eq{
{} [ \D_{iA} ,\D_{jB} ] = \delta_{ij} \Fddu{A}{B}{C} \D_{iC} .
}{DiDj}
Consequently, isometric invariants will in general be non-commutative as well.  

We also define the sum of isometric momenta of a set of particles labeled by $i\in I$,
\eq{
\D_{I} = \sum_{i\in I} \D_{i},
}
whose square $\D_I^2$ is the Casimir for that set of particles. 
From the commutation relation in \Eq{DiDj}, we trivially see that the Casimirs of a set $I$ and $J$ commute if those sets are disjoint or if one is contained in the other, so
\eq{
{} [\D^2_{I} ,\D^2_{J}] =  0 \quad \textrm{if} \quad I\supseteq J \textrm{ or } I\subseteq J \textrm{ or } I \cap J = \varnothing .
}{Casimir_comm}
Furthermore, we can compute the commutator of the total isometric momenta of a set $I$ with the Casimir on a set $J$,
\eq{
{} [\D_{I} ,\D^2_{J}] =  0 \quad \textrm{if} \quad I\supseteq J  \textrm{ or } I \cap J = \varnothing .
}{Casimir_D_comm}
Another useful commutator is 
\eq{
{} [\D_{I} , F_{ABC} \D_{i_1}^A \D_{i_2}^B \D_{i_3}^C] =  0 \quad \textrm{for} \quad i_1,i_2,i_3\in I,
}
by virtue of the Jacobi identity. More generally, $\D_I$ actually commutes with any isometric invariant differential operator $\mathbbm{f}$  constructed solely from isometric momenta of particles $i\in I$, so
\eq{
{} [\D_{I} , \mathbbm{f}(\D_{i_1},  \cdots,\D_{i_n})] =  0 \quad \textrm{for} \quad i_1,\cdots, i_n  \in I.
}{invar_commute}
We emphasize that $\mathbbm{f}$ is isometric invariant whenever all of its isometric indices are contracted.
This is the generalization of what happens for Lorentz invariants in flat space and conformal invariants in AdS.

In summary we have seen that the kinematics of particles in curved symmetric spacetimes can be recast in isometric frame in which the corresponding isometric momenta are literally isometry generators acting on the external kinematic data.  The isometric momenta behave precisely like flat space momenta except that they have non-commutative properties. 

\subsection{Isometric Representation}

We are now equipped to explore more complicated structures built up from contact correlators.  In particular, let us define a function $f$ that is equal to the action of a differential operator $\mathbbm{f}$ on the product wavefunction $\psi$, so 
\eq{
f(p_1,\cdots, p_n,x) =  \mathbbm{f}(\D_1, \cdots, \D_n) \psi(p_1,\cdots, p_n,x).
}{diff_rep}
Here we assume that $\mathbbm{f}$ is isometric invariant because its isometric indices are all contracted.   For obvious reasons, we refer to the differential operator $\mathbbm{f}$ as the isometric representation of $f$.  It is straightforward to 
show that $f$ is isometric invariant since it conserves isometric momentum,
\eq{
\left(\D+ \sum_{i=1}^n \D_{i} \right)f(p_1, \cdots, p_n,x) = 0.
}{f_annihilate}
Here the differential operator in parentheses annihilates $\psi$ due to \Eq{psi_annihilate} and commutes with $\mathbbm{f}$ due to \Eq{invar_commute}.   Trivially rearranging terms in \Eq{f_annihilate}, we find that
\eq{
\D f(p_1, \cdots, p_n,x) =  \left[ - \left( \sum_{i=1}^n \D_{i}\right) \mathbbm{f}(\D_1, \cdots, \D_n) \right] \psi(p_1,\cdots, p_n,x),
}{Df}
which says that $\D f$ also has an isometric representation defined by the differential operator in square brackets.  More generally, \Eq{f_annihilate} implies that any number of Lie derivatives of $f$ yields a function that also has an isometric representation. Note however, that since $\D f$ has an uncontracted isometric index, it is isometric covariant rather than invariant.

Furthermore, the product of functions that each have isometric representations always yields a  function that also has an isometric representation.  In particular, given a function $g$ for which
\eq{
g(p_{n+1}, \cdots, p_m,x) =  \mathbbm{g}(\D_{n+1}, \cdots, \D_m) \psi(p_{n+1},\cdots, p_m,x),
}{diff_rep_another}
then it is trivial to see that the product of $f$ and $g$ is 
\eq{
&f(p_1, \cdots, p_n,x) g(p_{n+1}, \cdots, p_m,x) = \big[ \mathbbm{f}(\D_1, \cdots, \D_n) \mathbbm{g}(\D_{n+1}, \cdots, \D_m)\big]  \psi(p_{1},\cdots, p_m,x),
}{fg} 
which is a new function which also has an isometric representation defined by the differential operator in square brackets.    Note that this assertion requires that $f$ and $g$ depend on disjoint sets of external kinematic labels, namely $p_1, \cdots, p_n$ and $p_{n+1}, \cdots, p_m$, respectively.  In all the expressions we will encounter in our analysis, this disjoint condition will be satisfied, and will in fact be mandated by locality.

In summary, \Eq{Df} and \Eq{fg} imply the following claim.   Starting from a collection of functions that each have an isometric representation and all depend on disjoint sets of external kinematic labels, any arbitrary combination of these functions and their Lie derivatives will also have an isometric representation.

\section{Correlators}

In this section we define a new class of observables---on-shell correlators---which can be computed on an arbitrary curved manifold.  
For symmetric manifolds, these calculations are drastically streamlined by the existence of isometric kinematics.   As we will see, on-shell correlators with isometric representations are literally identical in form to their flat space cousins, albeit with momenta replaced with isometric momenta that are intrinsically non-commutative.

\subsection{Definition}

The $n$-point on-shell correlator for scalars in curved spacetime is given by the formula,
\eq{
\Amp_n = \Amp(p_1, \cdots, p_n) = \left( \prod_{i=1}^n \int_{x_i} \psi(p_i,x_i) \nabla_{x_i}^2  \right) \langle \phi(x_1) \cdots \phi(x_n)\rangle
 \qquad \textrm{where} \qquad \nabla^2_{x_i}  \psi(p_i,x_i)=0,
}{Amp_def}
which is the generalization of LSZ reduction on the $n$-point position-space correlator.  In particular, the factors in parentheses amputate the external legs and dress them with external wavefunctions that satisfy the linearized equations of motion.

The on-shell correlator in \Eq{Amp_def} is a priori defined in any curved spacetime of arbitrary signature. Furthermore, it is the natural generalization of a number of commonly studied observables. In particular, in flat space it coincides with the on-shell scattering amplitude, while in AdS it corresponds to boundary correlators.  Our definition also includes transition amplitudes in Lorentzian AdS/CFT \cite{BalasubramanianGL99,Giddings99,Raju10} and the Lorentzian on-shell correlator that was defined for general spacetimes in \cite{MeltzerS20,Sivaramakrishnan21}.

Henceforth we will consider these on-shell correlators at leading order in perturbation theory, {\it i.e.}~tree level, and  for the case of symmetric spacetimes described in \Sec{sec:geo}.
Of course, it is certainly formally possible to compute the tree-level $n$-point correlator $\Amp_n$ using Feynman rules derived directly from the Lagrangian expressed in the usual coordinate basis.  In this case, $\Amp_n$ is equal to a sum of Feynman diagrams given by convolutions of products of the propagator, which is the inverse Laplacian $1/\nabla^2$, and the vertices, which are in general constructed from $\nabla$.

Nevertheless, as we will see shortly, it is substantially simpler to compute $\Amp_n$ after a transformation to isometric frame.  Crucially, this is only possible if the underlying spacetime manifold is symmetric, which we henceforth assume. 
 In this case, the propagator is the inverse Casimir $1/\D^2$ and the vertices are built from $\D$.  In this formalism, $\Amp_n$ can be recast in an isometric representation defined by a differential operator comprised of isometric momenta acting on the contact correlator, so
\eq{
\Amp(p_1, \cdots , p_n) = \AmpD(\D_1, \cdots, \D_n) \Cont(p_1, \cdots, p_n).
}{Amp_decomp}
While this procedure is mechanically straightforward, it involves propagators and vertices that are intrinsically non-commutative.  Naively, it is not clear precisely how to order these Feynman diagrammatic elements, which do not have an obvious ordering.  
As we will see shortly, however, it is actually more illuminating to compute $\Amp_n$ from the classical solutions to the equations of motion. This is possible due to the well-known fact that classical solutions are generating functionals of tree-level Feynman diagrams \cite{Boulware:1968zz,ramond1997field}. By employing this procedure, we will discover that the propagators and vertices are naturally ordered and there is in fact a natural separation between numerators and denominators in $\Amp_n$.

\subsection{Example: \texorpdfstring{$\phi^3$}{phi3} Theory}
\label{sec:phi3}

To best illustrate the mechanics of our formalism it will be useful to consider the simple case of massless scalar $\phi^3$ theory in a curved symmetric spacetime.  The Lagrangian and equations of motion for this theory are
\eq{
\L = \frac{1}{2} \nabla_\mu  \phi \nabla^\mu  \phi + \frac{\kappa}{6}\phi^3  \qandq
\nabla^2  \phi = \frac{\kappa}{2} \phi^2  .
}{phi3}
Transforming to the isometric representation, these expressions become
\eq{
\L = \frac{1}{2} \D_A  \phi \D^A  \phi + \frac{\kappa}{6}\phi^3  \qandq
\D^2  \phi = \frac{\kappa}{2} \phi^2  .
}{phi3_D}
It is now straightforward to perturbatively construct a solution to the equations of motion.  First, we initialize with a zeroth order solution $\phi(p_1, x)$ that depends on an arbitrary external kinematic label $p_1$. Second, we plug this zeroth order solution back into the equation of motion to then solve for the first order solution $\phi(p_1 ,p_2,x)$, which depends on $p_1$ and $p_2$.  The procedure is then iterated.
At each order, more and more wavefunctions are glued together via the equations of motion.  In the context of amplitudes, this algorithm is equivalent  to computing the Feynman diagrams via Berends-Giele recursion. 

Concretely, the general perturbative solutions for $\phi^3$ theory satisfy Berends-Giele recursion relations which can be graphically represented as
\begin{equation}
        \raisebox{-6pt}{
\begin{tikzpicture}
        \node[circle,draw, very thick, fill=white] (c) at (0,0){};
        \node[left] at (0,0) {\small $\phi(p_{k_1} , p_{k_2}, \cdots ,x)\, \,\,\,$};
\end{tikzpicture}}
\quad =\quad  \sum_{\substack{(I,J) = {\rm part}(K) }}
\raisebox{-46pt}{
\begin{tikzpicture}
    \draw [internalleg] (3,1.4) -- (1.5,0) ;
    \draw [internalleg] (3,-1.4)-- (1.5,0) ;
    \draw [internalleg] (1.5,0) -- (0,0)   ;
    \node[circle,draw, very thick, fill=white] (c) at (3,1.4){};
    \node[circle,draw, very thick, fill=white] (c) at (3,-1.4){};
    \node[right] at (3,1.4) {\small $\,\,\,\,\phi(p_{i_1} , p_{i_2}, \cdots  ,x)$};
    \node[right] at (3,-1.4) {\small $\, \,\,\,\phi(p_{j_1} , p_{j_2}, \cdots ,x)$};
\end{tikzpicture}}
\end{equation}
where the sum runs over partitions of the set $K=\{k_1, k_2, \cdots \}$ into non-empty subsets, $I = \{i_1, i_2, \cdots\}$ and $J=\{j_1, j_2, \cdots\}$.  Concretely, the perturbative solutions at low orders are
\eq{
\phi(p_1,p_2, x) &=  \frac{\kappa}{\D^2} \big[ \phi(p_1,x) \phi(p_2,x)  \big] \\
\phi(p_1,p_2, p_3,x) &= \frac{\kappa}{\D^2} \big[ \phi(p_1,x) \phi(p_2,p_3,x) +\phi(p_2,x) \phi(p_1,p_3,x) +\phi(p_3,x) \phi(p_1,p_2,x)  \big] \\
\phi(p_1,p_2, p_3,p_4, x) &= \frac{\kappa}{\D^2}  \big[\phi(p_1,x) \phi(p_2,p_3,p_4, x) +\cdots  +
\phi(p_1,p_2,x) \phi(p_3,p_4,x) +\cdots   \big],
}{phi3_recursion}
and similarly at higher points.  Note that the numerical factor of $1/2$ in \Eq{phi3_D} has been cancelled in \Eq{phi3_recursion} because the partitions $(I,J)$ and $(J,I)$ are treated as distinct and thus both contribute.
To solve for the on-shell correlators, we take the zeroth order seed for the solution to be a single external wavefunction,
\eq{
\phi(p_1, x) = \psi(p_1,x)   .
}{seed}
Plugging \Eq{seed} into \Eq{phi3_recursion}, we obtain the perturbative solutions
\eq{
\phi(p_1,p_2, x) &= \frac{\kappa}{\D_{12}^2}  \psi(p_1,p_2, x) \\
\phi(p_1,p_2, p_3,x) &= \frac{\kappa}{\D_{123}^2}   \left[  \frac{\kappa}{ \D_{23}^2}+ \frac{\kappa}{ \D_{13}^2}+ \frac{\kappa}{ \D_{12}^2}\right] \psi(p_1,p_2, p_3, x) \\
\phi(p_1,p_2, p_3,p_4, x) &= \frac{\kappa}{\D_{1234}^2}  \left[  \frac{\kappa}{\D_{234}^2}   \left[  \frac{\kappa}{ \D_{34}^2}+ \frac{\kappa}{ \D_{24}^2}+ \frac{\kappa}{ \D_{23}^2}\right] +\cdots +  \frac{\kappa^2}{\D_{12}^2 \D_{34}^2} +\cdots \right]\psi(p_1,p_2, p_3, p_4, x) .
}{phi3_recursion_sol}
Hence, we see that \Eq{phi3_recursion_sol} takes precisely the form of an isometric representation,
\eq{
\phi(p_1,\cdots, p_{n-1} ,x) &= \Phi(\D_1, \cdots, \D_{n-1}) \psi(p_1, \cdots, p_{n-1}, x).
}{phi_diff_rep}  
This is no accident---as described previously, products or Lie derivatives of functions with isometric representations will yield a new function that also has an isometric representation.  Since the equations of motion only involve products and Lie derivatives, the resulting perturbative solution inherits an isometric representation from the seed wavefunctions.  By inspection, we read off the corresponding differential operators from \Eq{phi3_recursion_sol}, 
\eq{
\Phi(\D_1,\D_2) &= \frac{\kappa}{\D_{12}^2} \\
\Phi(\D_1,\D_2, \D_3) &= \frac{\kappa}{\D_{123}^2} \left[ \frac{\kappa}{ \D_{12}^2} + \frac{\kappa}{ \D_{23}^2}+ \frac{\kappa}{ \D_{31}^2} \right] \\
\Phi(\D_1,\D_2, \D_3,\D_4)&= \frac{\kappa}{\D_{1234}^2} \left[  \frac{\kappa}{\D_{234}^2}   \left[  \frac{\kappa}{ \D_{23}^2}+ \frac{\kappa}{ \D_{34}^2}+ \frac{\kappa}{ \D_{42}^2}\right] +\cdots +  \frac{\kappa^2}{\D_{12}^2 \D_{34}^2} +\cdots \right].
}{phi3_Phi_sol}
Now plugging the perturbative solutions in \Eq{phi3_recursion_sol} into the definition of the on-shell correlator,
\eq{
\Amp(p_1,\cdots,p_n) &= \int_x \psi(p_n,x) \D^2 \phi(p_1,\cdots, p_{n-1} ,x) 
= \big[ \D_{1\cdots n}^2  \Phi(\D_1, \cdots, \D_{n-1}) \big] \Cont(p_1,\cdots, p_{n} ,x) ,
}
we obtain the isometric representation of the on-shell correlator as defined in \Eq{Amp_decomp}.  This object is simply related to that of the classical solution via
\eq{
\AmpD(\D_1, \cdots, \D_{n}) & = \D_{1\cdots n}^2  \Phi(\D_1, \cdots, \D_{n-1}),
}{amputate}
which corresponds to amputating the last leg. Plugging \Eq{phi3_Phi_sol} into the above equation, we arrive at our final result,
\eq{
\AmpD(\D_1,\D_2,\D_3) &= \kappa  \\
\AmpD(\D_1,\D_2, \D_3,\D_4) &=  \frac{\kappa^2}{ \D_{23}^2}+ \frac{\kappa^2}{ \D_{13}^2}+ \frac{\kappa^2}{ \D_{12}^2} \\
\AmpD(\D_1,\D_2, \D_3,\D_4,\D_5) &=     \frac{\kappa}{\D_{234}^2}   \left[  \frac{\kappa}{ \D_{34}^2}+ \frac{\kappa}{ \D_{24}^2}+ \frac{\kappa}{ \D_{23}^2}\right] +\cdots +  \frac{\kappa^2}{\D_{12}^2 \D_{34}^2} +\cdots  ,
}{phi3_Amp}
which defines the isometric representations for the on-shell correlators in $\phi^3$ theory.

\subsection{Example: \texorpdfstring{$\nabla\phi^4$}{dphi4} Theory}
\label{sec:dphi4}

The above analysis generalizes trivially to include higher-order potential terms and higher-derivative interactions.  To illustrate the latter,  consider a simple $\nabla \phi^4$ theory, whose Lagrangian and equations of motion are
\eq{
\L = \frac{1}{2} \nabla_\mu  \phi \nabla^\mu  \phi + \frac{\lambda}{8}( \nabla_\mu  \phi \nabla^\mu  \phi )^2  \qandq
\nabla^2  \phi = \frac{\lambda}{2} \nabla_\mu \left( \nabla^\mu \phi  \nabla_\nu \phi \nabla^\nu \phi\right)  .
}{phi4}
Again transforming to isometric frame, we obtain
\eq{
\L = \frac{1}{2} \D_A  \phi \D^A  \phi + \frac{\lambda}{8}(\D_A  \phi \D^A  \phi )^2  \qandq
\D^2  \phi = \frac{\lambda}{2} \D_A \left( \D^A \phi  \D_B \phi \D^B \phi\right)  .
}{phi4_D}
The equations of motion again imply Berends-Giele recursion relations whose perturbative solutions are given by
\eq{
\phi(p_1,p_2, p_3, x) &=  \frac{\lambda}{\D^2} \D_A \big[ \D^A \phi(p_1,x) \D_B \phi(p_2,x)\D^B \phi(p_3,x) +\cdots  \big] \\
\phi(p_1,p_2, p_3, p_4, p_5, x) &=  \frac{\lambda}{\D^2} \D_A \big[ \D^A \phi(p_1,x) \D_B \phi(p_2,x)\D^B \phi(p_3,p_4,p_5, x) +\cdots  \\
 &\qquad \quad \;\, +\D^A \phi(p_1, p_2, p_3,x) \D_B \phi(p_4,x)\D^B \phi(p_5, x) +\cdots  \big] ,
}{phi4_recursion}
and similarly for higher points.
Seeding the recursion with an external wavefunction as in \Eq{seed}, we obtain expressions for the perturbative solutions that have isometric representations as shown in \Eq{phi_diff_rep}, with the corresponding differential operators, 
\begin{align}\label{phi4_recursion_sol}
\Phi(\D_1,\D_2, \D_3,x) &= \frac{\lambda}{\D_{123}^2}   \left[ (\D_{123}\cdot \D_1)(\D_2 \cdot \D_3)  +\cdots \right] \\
\Phi(\D_1,
\cdots
\D_5,x) &= \frac{\lambda}{\D_{12345}^2}  \left[ (\D_{12345}\cdot \D_{1})(\D_2\cdot \D_{345}) \frac{\lambda}{\D_{345}^2}   \left[ (\D_{345}\cdot \D_3)(\D_4 \cdot \D_5) +\cdots \right] +\cdots \right. \nonumber \\
 &\qquad \quad \;\;\;  \left. +(\D_{12345}\cdot \D_{123}) \frac{\lambda}{\D_{123}^2}   \left[ (\D_{123}\cdot \D_1)(\D_2 \cdot \D_3) +\cdots \right](\D_4 \cdot \D_5)   +\cdots  \right] . 
 \nonumber 
 \end{align}
Again amputating these expressions via \Eq{amputate}, we obtain our final expressions,
\begin{align}
\label{phi4_Amp}
\AmpD(\D_1,\D_2,\D_3,\D_4) &= \lambda (\D_{123}\cdot \D_1)(\D_2 \cdot \D_3)  +\cdots
\\
\AmpD(\D_1,\D_2, \D_3,\D_4,\D_5,\D_6) &= \lambda  (\D_{12345}\cdot \D_{1})(\D_2\cdot \D_{345}) \frac{\lambda}{\D_{345}^2}   \left[ (\D_{345}\cdot \D_3)(\D_4 \cdot \D_5) +\cdots \right] +\cdots \nonumber  \\
 &  +\lambda (\D_{12345}\cdot \D_{123}) \frac{\lambda}{\D_{123}^2}   \left[ (\D_{123}\cdot \D_1)(\D_2 \cdot \D_3) +\cdots \right](\D_4 \cdot \D_5)   +\cdots  , \nonumber
 \end{align}
which are the isometric representations of the on-shell correlators in $\nabla \phi^4$ theory.

\subsection{General Prescription}
\label{sec:gen_pres}

Our analysis of $\phi^3$ theory and $\nabla\phi^4$ theory illustrate the mechanical steps by which one perturbatively solves the equations of motion to obtain an isometric representation for the on-shell correlators.  As is evident from the final expressions in \Eq{phi3_Amp} and \Eq{phi4_Amp}, however, one can compute these isometric representations more directly.

In particular, the approach of equations of motion and Berends-Giele recursion serves simply to define an ordering prescription for the isometric momenta, which are intrinsically non-commutative.  There is, however, an alternative but equivalent diagrammatic procedure which accomplishes the same goal.   To begin, recast the Lagrangian for the theory in question into isometric frame.  This is always possible in a broad class of theories, {\it e.g.}~for scalar theories with at most one derivative per field as described in \Eq{EOM_gen_D}.  Next, we apply the following algorithm:

\begin{itemize}

\item[{\it i})] Enumerate all tree-level Feynman diagrams contributing to the $n$-point on-shell correlator.   External legs are amputated while internal vertices and propagators are given by the Feynman rules in terms of isometric momenta.

\item[{\it ii})] Mark leg $n$ as the ``root'' and all remaining legs as ``leaves''.   The direction pointing towards the root will be denoted ``downwards'', while towards the leaves will be denoted ``upwards''.

\item[{\it iii})] Evaluate each Feynman diagram, starting from the leaf legs and sequentially multiplying vertices, propagators, or external wavefunctions that appear on the way downwards to the root leg.  The ordering of the isometric momenta is dictated by this order of appearance.

\item[{\it iv})] Make sure to write every vertex and propagator in terms of the isometric momentum variables that appear upwards from that graphical element.  This is equivalent to eliminating the isometric momentum of the root leg via total isometric momentum conservation.

\end{itemize}

\noindent

The above procedure is equivalent to an exceedingly simple procedure for lifting on-shell scattering amplitudes in flat space to on-shell correlators in curved spacetime.   First, compute the relevant flat space Feynman diagrams, taking special care to eliminate the momentum of the root leg. By locality, every Feynman diagram is a function of Lorentz invariants $p_I\cdot p_J$ in which the contracted momenta $p_I$ and $p_J$ are either disjoint, $I \cap J = \varnothing $, or contained in one another, so $ I\supseteq J $ or vice versa.  To lift to curved spacetime,  simply promote all flat space momenta $p_I$ to isometric momenta $\D_I$.  A priori, the relative ordering of any isometric momenta is ambiguous, but the prescription described above is equivalent to choosing  $\D_I$ to always be to the left of $\D_J$ when $ I\supseteq J $, and vice versa.
The application of this procedure for $\phi^3$ theory and $\nabla\phi^4$ theory will reproduce \Eq{phi3_Amp} and \Eq{phi4_Amp} directly. 

Step {\it iv}) appears rather innocuous but it actually has a very interesting implication for the structure of the on-shell correlator.  In particular, all propagators which appear in a given Feynman diagram actually commute with each other.  To see why, consider $\phi^3$ theory, where \Eq{phi3_Amp} exhibits strings of propagators such as
\eq{
\frac{1}{\D^2_{234}\D^2_{34}} \qandq \frac{1}{\D^2_{12}\D^2_{34}} .
}
Locality enforces that any pair of propagators that appear in a given Feynman diagram will either be nested or disjoint.  Thus, \Eq{Casimir_comm}, implies that all propagators that appear within a given Feynman diagram actually commute with each other.

Furthermore, the propagators not only commute with each other, but also with everything in the numerator that is downwards. In other words, in a symmetric curved space, locality is encoded not only in the propagator structure, but also in the structure of the numerators of Feynman diagrams. For example in $\nabla \phi^4$ theory, \Eq{phi4_Amp} exhibits propagators such as $1/\D^2_{345}$ and $1/\D^2_{123}$ which commute with everything downwards, {\it i.e.}~to the left of them.  This is possible because all isometric momenta which appear to the left of $1/\D^2_{345}$ either contain or are disjoint from that set of momenta, and similarly for $1/\D^2_{123}$.  Consequently, by \Eq{Casimir_D_comm} we can factor $1/\D^2_{345}$ and $1/\D^2_{123}$ out to the left.

This feature is quite general, and follows from the way in which on-shell correlators are computed from the equations of motion.   For example, in \Eq{phi3_recursion_sol} and \Eq{phi4_recursion_sol} it is clear that the propagators act from the left in each perturbative solution.  Since the propagator commutes with the total isometric momentum flowing through it, it will commute with any vertex into which that perturbative solution is fed.

Putting everything together, this implies that every on-shell correlator has a further decomposition into propagator denominators and numerators,
\eq{
\AmpD_n =\AmpD(\D_1,\cdots, \D_n) &= \sum_{\alpha } \frac{1}{\mathbbm{d}_\alpha(\D_1,\cdots, \D_n)} \times\mathbbm{n}_\alpha(\D_1,\cdots, \D_n) ,
}{num_denom}
where $\alpha$ runs over cubic topologies.   Crucially, all of the propagators in the denominator $\mathbbm{d}_\alpha$ are mutually commuting.  In contrast, the numerator $\mathbbm{n}_\alpha$ is a complicated non-commutative product of isometric momenta.  \Eq{num_denom} implies that there is a well-defined notion of numerator and denominator in curved spacetime even though the isometric momenta are intrinsically non-commutative.

Notably, the decomposition of amplitudes into numerators and denominators is central to color-kinematics duality and the application of the double copy.   This suggests the possibility that these structures may have curved spacetime generalizations, offering a new and in principle highly efficient approach to computing on-shell correlators.  As we will see, this is indeed the case for certain theories.

\subsection{Field Basis Invariance}
\label{sec:field_basis}

As is well-known, on-shell scattering amplitudes in flat space are invariant under redefinitions of the field basis.  Remarkably, the same is true of on-shell correlators in curved spacetime.
For concreteness, let us return to the example of $\phi^3$ theory discussed earlier.  We apply the following field redefinition to the scalar,
\eq{
\phi(x) \quad  \rightarrow \quad \phi'(x) = \phi(x) + \tau \phi(x)^2 +\cdots,
}
where the ellipses denote higher order nonlinearities in the transformation.  
After this field redefinition, the off-shell Feynman rules will change.  
Meanwhile, the on-shell correlators map to
\eq{
A(p_1,\cdots,p_n) \quad \rightarrow\quad A'(p_1,\cdots,p_n),
}
where the transformed on-shell correlator is
\eq{
\Amp'(p_1,\cdots,p_n) &= \int_x \psi(p_n,x) \D^2 \big[ \phi(p_1,\cdots, p_{n-1} ,x) + \tau\left( \phi(p_1, x) \phi(p_2,\cdots, p_{n-1} ,x)   +\cdots \right)\big],
}
which is obtained by applying the field redefinition to the Berends-Giele recursion relations. The parameter $\tau$ flags all the new contributions arising from the field redefinition,  which involve all partitions of legs $1, \cdots, n-1$ into the pair of scalar fields in the transformation.    After plugging in the perturbative solutions for the scalar fields, we obtain
\eq{
\Amp'(p_1,\cdots,p_n) &= \int_x \psi(p_n,x) \D_{1 \cdots n-1}^2 \left[\frac{1}{\D_{1 \cdots n-1}^2}(\cdots\!) +\frac{\tau}{ \D_{2 \cdots n-1}^2 }(\cdots\!)+ \cdots \right]\psi(p_1, \cdots, p_{n-1},x),
}
where the propagators shown explicitly all arise from the root legs of the scalar classical field solutions.  

Next, we show how the isometric on-shell condition for the $n$-th leg actually truncates all nonlinear contributions from the field redefinition.  An important subtlety arises from total isometric momentum conservation, which implies that the isometric momenta are only defined up to the overall constraint, $\D_1 + \cdots + \D_n=0$.  In order to impose the on-shell condition on the $n$-th leg, we have to transform to a basis of independent isometric momenta.  This is achieved by eliminating the momentum of the $(n-1)$-th leg from the correlator,
\eq{
\Amp'(p_1,\cdots,p_n) &= \int_x \D_{n}^2 \left[\frac{1}{\D_{n}^2}(\cdots\!) +\frac{\tau}{ \D_{1n}^2 }(\cdots\!)+ \cdots \right]\psi(p_1, \cdots, p_{n},x) \\
& =  \left[(\cdots\!) +\frac{\tau \D_n^2 }{ \D_{1n}^2 }(\cdots\!)+ \cdots \right]\Delta(p_1, \cdots, p_{n}) = A(p_1,\cdots, p_n).
}
In the final equality we have used that $\D_n^2$ vanishes, thus eliminating all terms proportional to $\tau$, which were new contributions induced by the field redefinition.  

In summary, we have proven that the isometric on-shell condition for the $n$-th leg enforces that the on-shell correlator is invariant under changes of field basis, just as for on-shell scattering amplitudes in flat space.  As should be apparent, the argument we have outlined closely mirrors the usual flat space proof.

\section{Color-Kinematics Duality}

In this section we derive the duality between color and kinematics for BAS theory and NLSM in curved spacetime while in isometric frame.   By generalizing the approach of \cite{CCK} to curved backgrounds, we show that BAS is dual to NLSM under the mapping of the algebra of color to the algebra of gauged isometries.

\subsection{Biadjoint Scalar Theory}

As is well-known, the template for all theories with color-kinematics duality is BAS theory.  We now we review this theory briefly, taking note of its symmetries and conserved charges.  We then rewrite these structures in isometric frame.

\subsubsection{Equations of Motion} 

The scalar field $\phi^{a \ov a}$ of BAS theory is an adjoint of a color and dual color symmetry, which are labeled by unbarred and barred indices $a,b,c$, etc., and $\ov a, \ov b, \ov c$, etc., respectively.  To eliminate notational clutter, we repackage the BAS field as a color matrix,
\eq{
\phi^{\ov a} = \phi^{a \ov a} t_{a},
}
while the dual color index is left free and uncontracted.  In this notation, the equation of motion of BAS theory in curved spacetime is
\eq{
\nabla^2 \phi^{\ov c}  +  \frac{1}{2}  \fbarddu{ a}{ b}{ c}  [\phi^{\ov a} , \phi^{\ov b} ] = 0 .
}
 Under the dual color symmetry, the fields transform as
\eq{
\phi^{\ov c} \rightarrow  \phi^{\ov c}  +   \fbarddu{ a}{ b}{ c} \theta^{\ov a} \phi^{\ov b},
}
and there is an associated dual color current,
\eq{
{\cal K}^{\ov c}_\mu =  \fbarddu{ a}{ b}{ c} \tr(\phi^{\ov a} \lr{\nabla}{\mu} \phi^{ \ov b}),
}
where the trace runs over the suppressed color indices.
On the support of the equations of motion, this current is conserved,
\eq{
\nabla^\mu {\cal K}^{\ov c}_\mu =  \fbarddu{ a}{ b}{ c}  \tr(\phi^{ \ov a} \lrsq{\nabla} \phi^{ \ov b} )= \fbarddu{ a}{ b}{ c}    \fbarddu{ d}{ e}{ a} \tr(\phi^{ \ov b} [ \phi^{ \ov d},\phi^{\ov e}] ) =0,
}
where we have used the Jacobi identity for dual color in the final step.

\subsubsection{Isometric Representation} It is trivial to the lift the BAS equations of motion into isometric frame.  Using \Eq{Laplacian_Lie}, we obtain the equations of motion 
\eq{
\D^2 \phi^{\ov c}  +  \frac{1}{2}  \fbarddu{ a}{ b}{ c}  [\phi^{\ov a} , \phi^{\ov b} ] = 0 .
}{EOM_BAS_iso}
Meanwhile, in isometric frame the dual color current is
\eq{
{\cal K}^{\ov c}_\mu =  \fbarddu{ a}{ b}{ c} \tr(\phi^{\ov a} \lr{\D}{\mu} \phi^{ \ov b}),
}{K_BAS_iso}
which, just as before, is conserved on the equations of motion,
\eq{
\nabla^\mu {\cal K}^{\ov c}_\mu =  \fbarddu{ a}{ b}{ c}  \tr(\phi^{ \ov a} \lrsq{\D} \phi^{ \ov b} )= \fbarddu{ a}{ b}{ c}    \fbarddu{ d}{ e}{ a} \tr(\phi^{ \ov b} [ \phi^{ \ov d},\phi^{\ov e}] ) =0,
}
where we have again used the dual color Jacobi identity.

\subsubsection{Correlators}

The on-shell correlators for BAS theory are identical to those of $\phi^3$ theory, except with each diagram dressed with color and dual color factors.  Applying the same procedure outlined earlier, we find that the isometric representations for these on-shell correlators are
\eq{
\AmpD(\D_1,\D_2,\D_3) = &\phantom{{}+{}}   f_{a_1 a_2 a_3} f_{\ov a_1 \ov a_2 \ov a_3} \\
\AmpD(\D_1,\D_2, \D_3,\D_4) = &\phantom{{}+{}}     \fddu{a_1}{a_2}{b} f_{ba_3 a_4}  \fddu{\ov a_1}{\ov a_2}{b} f_{\ov b \ov a_3 \ov a_4} \frac{1}{ \D_{12}^2}  
 + \textrm{2 terms} \\
\AmpD(\D_1, \D_2, \D_3, \D_4, \D_5)  =&   \phantom{{}+{}}    \fddu{a_1}{a_2}{b_1}  \fddu{b_1}{a_3}{b_2}  \fddu{b_2}{a_4}{a_5}   \fbarddu{a_1}{a_2}{b_1} \fbarddu{b_1}{a_3}{b_2}  \fbarddu{b_2}{a_4}{a_5}    \frac{1}{\D_{123}^2 \D^2_{12}}   + \textrm{11 terms} \\
&+    \fddu{a_1}{a_2}{b_1}  \fddu{b_2}{b_1}{a_5} \fddu{a_3}{a_4}{b_2}   \fbarddu{a_1}{a_2}{b_1}  \fbarddu{b_2}{b_1}{a_5}\fbarddu{a_3}{a_4}{b_2}   \frac{1}{\D_{12}^2 \D_{34}^2}  + \textrm{2 terms},
}{BAS_Amp}
and so on and so forth.  Since the propagator denominators are all mutually commuting, these isometric representations are literally identical to the expressions for on-shell scattering amplitudes in flat space computed via Feynman diagrams.

\subsection{Nonlinear Sigma Model}

Next, we formulate the NLSM in curved spacetime.  To do so we simply take the first-order formulation of the NLSM which manifests color-kinematics duality in flat space \cite{CCK} and lift it to curved spacetime.   

\subsubsection{Equations of Motion } The NLSM can be described by a first-order formulation in which the dynamical degree of freedom is the chiral current,
\eq{
j^{\mu} = j^{a\mu} t_a,
} which is an adjoint of color.  We impose that the chiral current is pure gauge and has vanishing divergence, so
\eq{
\nabla^{[\mu } j^{\nu]}  + [ j^{\mu}, j^{\nu}] = 0 \qandq \nabla_\mu j^\mu=  0.
}{EOM1st_NLSM}
As described at length in \cite{CCK} the above equations are equivalent to the more standard formulation of the NLSM.
Summing various derivatives of the expressions in \Eq{EOM1st_NLSM}, we obtain a second-order equation of motion
\eq{
\nabla^2 j^\mu  + \nabla^{[\mu} \nabla^{\nu]} j_{\nu} + [  j^{\nu} , \nabla_\nu j^{  \mu}]  = 0 ,
}{EOM_NLSM}
which governs the evolution of the chiral current.

\subsubsection{Isometric Representation}
Next, we transform the equations of motion in \Eq{EOM_NLSM} to isometric frame.  To begin, let us define the chiral current in isometric frame,
\eq{
j^A =  j^\mu K^A_\mu .
}
In terms of this expression, the linearized field strength for the chiral current is
\eq{
 \K_\mu^A \K_\nu^B \nabla^{[\mu } j^{\nu]} &=
  \K_\nu^B (\D^A j^\nu + \nabla_\mu K^{A\nu} j^\mu) -  \K_\mu^A (\D^B j^\mu + \nabla_\nu K^{B\mu} j^\nu)  
= \D^{[A } j^{B]} -  F^{A B C}  j_{C} ,
}
where we have used the Killing equation in \Eq{Killing_equation} and the commutator relation in \Eq{Killing_algebra_extra}.
Meanwhile, the divergence of the chiral current is
\eq{
\nabla_\mu j^\mu  & = g_{AB}  \nabla^\nu  (K_{\mu}^A K_{\nu}^B  j^\mu) 
= g_{AB} K_{\nu}^B \nabla^\nu (  K_{\mu}^A j^\mu ) = \D_A j^A.
}
Plugging the above expressions back into the NLSM equations of motion in \Eq{EOM1st_NLSM}, we obtain
\eq{
\D^{[A } j^{B]} -   F^{ABC}  j_{C}+  [j^{A} , j^{B}] =0 \qandq \D_A j^A=0.
}{EOM1st_NLSM_iso_comp}
Combining various Lie derivatives of these equations, we obtain the equation of motion for the NLSM in the isometric representation,
\eq{
\D^2 j^C  + [ j^{A},  \D_A  j^{C}] = 0 ,
}{EOM_NLSM_iso}
which is exactly the same as in flat space but with partial derivatives replaced with Lie derivatives.

\subsubsection{Correlators}

Following the procedure described for $\phi^3$ theory and $\nabla\phi^4$ described in \Sec{sec:phi3} and \Sec{sec:dphi4}, we can now compute the on-shell correlators of the NLSM.  A priori, this calculation can be performed starting from any particular formulation of the NLSM.  However, with color-kinematics duality in mind, we will opt for the NLSM equations of motion defined in \Eq{EOM_NLSM_iso},
\eq{
\D^2 j^{cC} + \fddu{a}{b}{c} j^{aA} \D_A j^{bC} =0,
}{EOM_NLSM_iso_comp}
shown here with all color indices written out explicitly.  Similar to before, these equations of motion imply Berends-Giele recursion relations for the chiral current, whose perturbative solutions are
\eq{
j^{cC}(p_1,p_2,x) &= -\frac{\fddu{a}{b}{c}}{\D^2} \big[j^{aA}(p_1,x) \D_A j^{bC}(p_2,x) + j^{aA}(p_2,x) \D_A j^{bC}(p_1,x)  \big] \\
j^{cC}(p_1,p_2,p_3, x) &= -\frac{\fddu{a}{b}{c}}{\D^2} \big[ j^{aA}(p_1,x) \D_A j^{bC}(p_2,p_3,x) +  j^{aA}(p_2,x) \D_A j^{bC}(p_1,p_3,x)+ \cdots \\
& \quad \quad \,\,\;\; \; + j^{aA}(p_1,p_2,x) \D_A j^{bC}(p_3,x)+ j^{aA}(p_1,p_3,x) \D_A j^{bC}(p_2,x) +\cdots\big].
}{BG_NLSM}
It is straightforward to perturbatively solve these equations provided we clarify two important subtleties.  First of all, what is the seed for the recursion?  Second, how do we extract the on-shell correlator for the scalar field $\pi^a$ of the NLSM from that of the chiral current $j^{aA}$, which is of course not precisely the same?  Both of these questions are answered by constructing the explicit map between $\pi^a$ and $j^{aA}$, as described in detail for flat space in \cite{CCK}.

In particular, we recall that the pure gauge condition in \Eq{EOM1st_NLSM} is imposed precisely so that the chiral current takes the form 
\eq{
j^A = i g^{-1} \D^A g,
}
for a color adjoint field $g$.  The NLSM scalar is embedded within $g$ via 
\eq{
g = 1- i\pi +\cdots,
}
where $\pi = \pi^a t_a$ and the ellipses denote nonlinear terms in the NLSM field that depend on the precise field basis.  In terms of the NLSM scalar, the chiral current is then
\eq{
j^{aA} = \D^A \pi^a + \cdots,
}{j_from_pi}
where the ellipses are again nonlinear in fields.  Consequently, we can extract $\pi^a$ from $j^{aA}$, up to nonlinear terms, by dotting the latter into a peculiar polarization vector
\eq{
\pi^a = \widetilde{\varepsilon}_A j^{aA}+ \cdots \quad \textrm{where} \quad  \widetilde{\varepsilon}_A = \frac{q_A}{q \cdot \D},
}{pi_from_j}
where $q_A$ is an arbitrary reference vector.

From \Eq{j_from_pi} we can deduce the seed for the Berends-Giele recursion.  By definition, the seed is a solution to the linearized equations of motion, so the terms in ellipses can all be dropped.  Let us denote the color-dressed on-shell wavefunction of the NLSM scalar by
\eq{
\pi^a(p_1,x) = \psi^a(p_1,x),
}
as well as the color-dressed product wavefunction and contact correlator,
\eq{
\psi^{a_1\cdots a_{n-1}}(p_1,\cdots, p_{n-1}, x) &= \prod_{i=1}^{n-1} \psi^{a_i}(p_i,x)\\
\Delta^{a_1\cdots a_{n-1}}(p_1,\cdots, p_{n-1}) &= \int_x \psi^{a_1\cdots a_{n-1}}(p_1,\cdots, p_{n-1}, x).
}
Here the free color indices correspond to the external color polarizations of the wavefunctions.
  Consequently, \Eq{j_from_pi} implies that
\eq{
j^{aA}(p_1,x) =  \D^A \pi^a(p_1,x) = \D^A \psi^a(p_1,x) ,
}{seed_NLSM}
so the on-shell wavefunction for the chiral current is equal to an isometry-generating Lie derivative acting on the on-shell wavefunction for the NLSM scalar.\footnote{In the context of AdS, this implies that the bulk-boundary propagators for the chiral current and NLSM scalar are related by a conformal transformation, $\D_A$.  Since $\D_A$ is a scalar, dimensionless operator that commutes with the conformal Casimir, $\D^2$, these bulk-boundary propagators correspond to boundary operators with the same spin and conformal dimension.  }  

Moreover, \Eq{pi_from_j} implies that the $n$-point classical solution for the NLSM scalar is 
\eq{
\pi^a(p_1,\cdots, p_{n-1},x) = \widetilde{\varepsilon}_A j^{aA}(p_1,\cdots, p_{n-1},x) + \cdots,
}{final_dot_NLSM}
where the ellipses are nonlinear terms that depend on the field basis.   However, as described in \Sec{sec:field_basis}, these nonlinear terms will vanish identically when the root leg is taken to be on-shell.

We also define the isometric representation for the NLSM scalar,
\eq{
\pi^a(p_1,\cdots, p_{n-1},x) =  {\mathbb{\Pi}}^{a_1\cdots a_{n-1} a} (\D_1,\cdots, \D_{n-1}) \psi_{a_1 \cdots a_{n-1}} (p_1, \cdots, p_{n-1}, x),
}{pi_from_Pi}
which is related to the on-shell correlator for NLSM scalars via
\eq{
\Amp(p_1,\cdots,p_n) &= \int_x \psi^{a_n}(p_n,x) \D^2 \pi_{a_n}(p_1,\cdots, p_{n-1} ,x) \\
&= \big[ \D_{1\cdots n}^2  {\mathbb{\Pi}}_{a_1\cdots a_{n}} (\D_1, \cdots, \D_{n-1}) \big] \Cont^{a_1\cdots a_n}(p_1,\cdots, p_{n} ) . 
}
The corresponding isometric representation is then
\eq{
\AmpD_{a_1 \cdots a_n}(\D_1,\cdots,\D_n) &= \D_{1\cdots n}^2  {\mathbb{\Pi}}_{a_1\cdots a_{n}} (\D_1, \cdots, \D_{n-1}) ,
}{A_from_Pi}
which is just the amputation of the isometric representation of the NLSM scalar solution of the Berends-Giele recursion relations.

Putting it all together, the isometric representation for the on-shell correlator for NLSM scalars is computed as follows.  First, we compute the classical solution for $j^{aA}$ by solving the Berends-Giele recursion relations in \Eq{BG_NLSM} with the seed wavefunction in \Eq{seed_NLSM}.  Next, we contract the root leg with polarization $\widetilde{\varepsilon}_A$ as in  \Eq{final_dot_NLSM} to map the chiral current to the NLSM scalar.  Finally, we extract the isometric representation of the NLSM scalar solution as defined in \Eq{pi_from_Pi} and compute the on-shell correlator by amputating the root leg as in \Eq{A_from_Pi}.

As discussed in \Sec{sec:gen_pres}, however, solving the Berends-Giele recursion relations is identical to computing the standard Feynman diagrams in the theory while carefully keeping track of the ordering of isometric momenta. Consequently, the procedure we have outlined above is identical to implementing the following Feynman rules, 
\eq{
\raisebox{2pt}{
\begin{tikzpicture}
    \draw [internalleg](3,0) -- (0,0);
    \node[left] at (0,0) {\small $I$};
    \node[right] at (3,0) {\small $J$};
\end{tikzpicture}}
& \quad = \quad \frac{\delta_{a_I a_J} g_{A_I A_J}}{\D_I^2}
\\
\raisebox{-38pt}{
\begin{tikzpicture} 
    \draw [internalleg] (3,1.4) -- (1.5,0) ;
    \draw [internalleg] (3,-1.4)-- (1.5,0) ;
    \draw [internalleg] (1.5,0) -- (0,0)   ;
    \node[left] at (0,0) {\small $K$};
    \node[right, above] at (3,1.4) {\small $I$};
    \node[right, below] at (3,-1.4) {\small $J$};
\end{tikzpicture}} & \quad =\quad  \fddu{a_I}{a_J}{a_K} \left[(\D_I)_{A_J}  \delta^{A_K}{}_{A_I} - (\D_J)_{A_I}  \delta^{A_K}{}_{A_J}  \right]
\\
\raisebox{0pt}{
\begin{tikzpicture}
    \draw[leafleg] (3,0)--(0,0);
    \node[left] at (0,0) {\small $I$};
    \node[right] at (3,0) {\small $\hphantom{I}$};
\end{tikzpicture}}       & \quad =\quad (\D_I)^{A_I}  \\
\raisebox{0pt}{
\begin{tikzpicture}
    \draw[rootleg](3,0) -- (0,0);
    \node[right] at (3,0) {\small $I$};
\end{tikzpicture}}   & \quad =\quad \frac{q_{A_I}}{ q \cdot \D_{I}}  .
}{NLSM_Feynman_Rules}
Here the interaction vertex in the second line above is defined from the equations of motion in \Eq{EOM_NLSM_iso_comp}, while the leaf and root leg polarizations in the third and fourth lines are defined by \Eq{seed_NLSM} and \Eq{final_dot_NLSM}.  As described earlier, the ordering of isometric momenta is dictated by the relative location of graphical elements, {\it i.e.}~whether they are further upwards or downwards. 

For example, the isometric representation of the three-point on-shell correlator is  
\eq{
\AmpD_{a_1 a_2 a_3}(\D_1,\D_2,\D_3)&=
\raisebox{-37pt}{
\begin{tikzpicture}[scale=0.6]
    \draw [leafleg] (3,1.4) -- (1.5,0) ;
    \draw [leafleg] (3,-1.4)-- (1.5,0) ;
    \draw [rootleg] (1.5,0) -- (0,0)   ;
    \node[left] at (0,0) {\small $3$};
    \node[right, above] at (3,1.4) {\small $1$};
    \node[right, below] at (3,-1.4) {\small $2$};
\end{tikzpicture} 
}
                                 =  f_{a_1 a_2 a_3} \frac{q_A}{q \cdot \D_3} \left[  \D_2^B \D_{1B} \D_1^A- \D_1^B  \D_{2B} \D_2^A \right]\\
&=  f_{a_1 a_2 a_3} \frac{q_A}{q \cdot \D_3} (\D_1 \cdot \D_2) ( \D_1^A- \D_2^A )= 0,
}
where we have used that $\D_1\cdot \D_2 \sim \D_3^2 =0$ for three-point on-shell kinematics.
Meanwhile, for the four-point on-shell correlator we obtain
\eq{
&\AmpD_{a_1 a_2 a_3 a_4}(\D_1,\D_2,\D_3,\D_4) =
\raisebox{-42pt}{
\begin{tikzpicture}[scale=0.5]
    \draw [leaflegs] (3.75,2.1) -- (2.625,+1.05) ;
    \draw [leafleg] (3.75,-2.1)-- (1.5,0) ;
    \draw [internalleg] (2.625,+1.05) --  (1.5,0) ;
    \draw [leaflegs] (3.75,0)   -- (2.625,+1.05) ;
    \draw [rootleg] (1.5,0) -- (0,0)   ;
    \node[left] at (0,0) {\small $4$};
    \node[right, above] at (3.75,2.1) {\small $1$};
    \node[right, below] at (3.75,0) {\small $2$};
    \node[right, below] at (3.75,-2.1) {\small $3$};
\end{tikzpicture} 
} + 
\raisebox{-42pt}{
\begin{tikzpicture}[scale=0.5]
    \draw [leafleg] (3.75,2.1) -- (1.5,0) ;
    \draw [leaflegs] (3.75,-2.1)-- (2.625,-1.05) ;
    \draw [internalleg] (2.625,-1.05) --  (1.5,0) ;
    \draw [leaflegs] (3.75,0)   -- (2.625,-1.05) ;
    \draw [rootleg] (1.5,0) -- (0,0)   ;
    \node[left] at (0,0) {\small $4$};
    \node[right, above] at (3.75,2.1) {\small $1$};
    \node[right, above] at (3.75,0) {\small $2$};
    \node[right, below] at (3.75,-2.1) {\small $3$};
\end{tikzpicture} 
}
+ 
\raisebox{-42pt}{
\begin{tikzpicture}[scale=0.5]
    \draw [leaflegs] (3.75,2.1) -- (2.625,+1.05) ;
    \draw [leafleg] (3.75,-2.1)-- (1.5,0) ;
    \draw [internalleg] (2.625,+1.05) --  (1.5,0) ;
    \draw [leaflegs] (3.75,0)   -- (2.625,+1.05) ;
    \draw [rootleg] (1.5,0) -- (0,0)   ;
    \node[left] at (0,0) {\small $4$};
    \node[right, above] at (3.75,2.1) {\small $1$};
    \node[right, below] at (3.75,0) {\small $3$};
    \node[right, below] at (3.75,-2.1) {\small $2$};
\end{tikzpicture} 
}
                                 \\
&=  \fddu{a_1}{a_2}{b}f_{b a_3 a_4} \frac{q_A}{q \cdot \D_4} \!\left[ \D_3^B  \D_{12B} \Big( \frac{1}{\D_{12}^2 }(\D_1\cdot \D_2) (\D_1^A \!- \D_2^A)  \Big) \! - \!\Big( \frac{1}{\D_{12}^2 }(\D_1\cdot \D_2) (\D_1^B\!- \D_2^B)  \Big)   \D_{3B} \D_3^A \right] \\
& \phantom{=} + \textrm{2 terms} \\
&= \frac{1}{2}  \fddu{a_1}{a_2}{b}f_{b a_3 a_4} \frac{1}{q \cdot \D_4} \left[(\D_{3}\cdot \D_{12}) (q\cdot \D_1- q\cdot \D_2)- (\D_{1}\cdot \D_3-\D_{2}\cdot \D_3) (q\cdot \D_3) \right] +\textrm{2 terms}\\
&= \frac{1}{2}  \left[ \fddu{a_1}{a_2}{b}f_{b a_3 a_4} \D_1\cdot \D_3+ \fddu{a_1}{a_3}{b}f_{b a_2 a_4} \D_1\cdot \D_2\right],
}{NLSM_amps}
 where in the second equality we have grouped terms in parentheses so their origin from the equations of motion is clear.  Furthermore, in the last line we have transformed to a minimal DDM basis for color and also used that the $1/\D_4$ factor from the root polarization commutes with all isometric invariants, as per \Eq{invar_commute}.  This fact is generally true for any $n$-point on-shell correlator.  As expected, the dependence on the reference $q$ has dropped out of \Eq{NLSM_amps}, and the resulting color-stripped four-point on-shell correlator, $\sim \D_1\cdot \D_3$, is correct. 

We have also computed the six- and eight-point on-shell correlators for the NLSM using the above color-kinematic dual procedure.  The former precisely agrees with the result of \cite{DiwakarHRT21}, which is a nontrivial check that our formulation of the NLSM is equivalent to the standard one.  Meanwhile, the latter expression is a genuinely new result.   

Perhaps surprisingly, the curved spacetime NLSM on-shell correlators computed via this method---namely, using the color-kinematic dual equations of motion in \Eq{EOM_NLSM_iso_comp}---are literally identical to the corresponding on-shell scattering amplitudes in flat space.  In particular, the expressions are the same up to a trivial replacement of momenta with isometric momenta, as is evident from the very last line of \Eq{NLSM_amps}.  This occurs because the resulting on-shell correlators are a function of Killing Casimirs that mutually commute.  In particular, every term is given by a numerator and denominator composed of invariants $\D_I^2$ and $\D_J^2$ where the sets are either disjoint, $I \cap J = \varnothing $, or contained in one another, so $ I\supseteq J $ or vice versa.  The reason this happens is that in a two-derivative theory all interaction vertices are of the form $\D_I \cdot \D_J = \frac{1}{2}( \D_{IJ}^2- \D_I^2 - \D_J^2)$, which is a set of commuting Killing Casimirs.\footnote{We thank the authors of \cite{HRTpreprint} for pointing out this simple argument to us.}

\subsection{Kinematic Algebra}

The equations of motion for BAS theory and NLSM in \Eq{EOM_BAS_iso} and \Eq{EOM_NLSM_iso} are related to each other by color-kinematics duality at the level of fields, just as in flat space \cite{CCK}.  
By directly comparing these equations, it is straightforward to derive the precise mapping between them.  Furthermore, we will find that the on-shell correlators computed in our formulation of the NLSM have kinematic numerators that automatically satisfy the kinematic Jacobi identities.

\subsubsection{Field Theoretic Formulation}

Remarkably, BAS theory maps to the NLSM under a  literal replacement of dual color generators with isometry-generating Lie derivatives,
\eq{
t_{\ov a}  \quad \rightarrow \quad \D_A.
}
This substitution sends any dual colored field to a vector field, so for example 
\eq{
{\cal V}^{\ov a} \quad \rightarrow\quad {\cal V}^A,
}{sub_field}
where all color indices are left untouched.  Similarly, the dual color structure constants of BAS theory map to 
\eq{
\fbarddu{a}{b}{c}  {\cal V}^{\ov a}  {\cal W}^{\ov b} \quad\rightarrow\quad {\cal V}^{A}\D_A  {\cal W}^{C} - {\cal W}^{A}\D_A  {\cal V}^{C} ,
}{sub_struct}
which define the kinematic structure constants of the NLSM.

By applying the replacement rules in \Eq{sub_field} and \Eq{sub_struct}, we directly map the equation of motion for BAS theory in  \Eq{EOM_BAS_iso} to that of the NLSM in \Eq{EOM_NLSM_iso}.  As it turns out, this substitution has an elegant physical interpretation.  First, recall that the dual color algebra is obtained by the commutator,
\eq{
{} [{\cal V}^{\ov a}t_{\ov a},  {\cal W}^{\ov b} t_{\ov b}] = {\cal V}^{\ov a}  {\cal W}^{\ov b} \fbarddu{a}{b}{c} t_{\ov c}.
}{VW_color}
Hence, the dual color structure constants are obtained by commuting a pair of spacetime-dependent color generators parameterized by ${\cal V}^{\ov a}$ and ${\cal W}^{\ov b}$.  
Second, consider the commutator 
\eq{
{} [{\cal V}^{A}\D_{A},  {\cal W}^{B} \D_B] = \left({\cal V}^{A}\D_A  {\cal W}^{C} - {\cal W}^{A}\D_A  {\cal V}^{C} +  {\cal V}^{A} {\cal W}^{B} \Fddu{A}{B}{C} \right) \D_C,
}{VW_kinematic}
where the last term encodes the usual Killing algebra and the first and second terms arise from the fact that ${\cal V}^{A}$ and ${\cal W}^{B}$ are not constant.
As long as we work purely in isometric frame, all tensors will have isometric indices rather than metric indices.  Consequently, every Lie derivative acts on a scalar and so the very last term on the right-hand side of \Eq{VW_kinematic} will always ultimately vanish by \Eq{FKKD}.     Thus, we discover that the commutator in \Eq{VW_kinematic} precisely reproduces the kinematic structure constant in the replacement rule in \Eq{sub_struct}.  In other words, the kinematic structure constants are obtained by commuting a pair of spacetime-dependent, isometry-generating Lie derivatives parameterized by ${\cal V}^{A}$ and ${\cal W}^{B}$.  

In conclusion, we have learned that on a symmetric spacetime manifold, BAS theory maps to the NLSM under a duality between color and kinematics that sends
\eq{
\begin{array}{c}
\textrm{algebra of } \\
\textrm{dual color} 
\end{array} \quad \rightarrow \quad
\begin{array}{c}
\textrm{algebra of } \\
\textrm{gauged isometries}  .
\end{array}
}
Here we have introduced the notion of a gauged isometry because the field-dependent Lie derivatives in \Eq{VW_kinematic} describe precisely such a transformation.  This result is the curved space generalization of the flat space result of \cite{CCK}, where the kinematic algebra of the NLSM found to literally be the algebra of diffeomorphisms, also known as gauged translations.    A similar duality and double copy was also observed in certain fluid dynamical theories in \cite{Cheung:2020djz}.

Note also that the algebra of gauged isometries is actually distinct from the algebra of global isometries, {\it i.e.}~the Killing algebra, which corresponds to just the last term of \Eq{VW_kinematic}.  For the special case of flat space, this corresponds to the fact that the algebra of diffeomorphisms is distinct from the algebra of translations, which is trivial.

Armed with an understanding of color-kinematics duality at the level of fields, we can now apply the replacement rules in \Eq{sub_field} and \Eq{sub_struct} to map any quantity in BAS theory to an equivalent one in the NLSM.  Under this mapping, the dual color current for BAS theory in \Eq{K_BAS_iso} becomes the kinematic current of the NLSM,
\eq{
 {\cal K}_{A}^C =   \tr ( j^{ B } \lr{\D}{A}  \D_B j^C ),
}
whose divergence can be computed explicitly to be
\eq{
\D^A {\cal K}_{A}^C & =  \tr(  j^{ B } \lrsq{\D}  \D_B j^C  )= -\tr(  j^{B } \D_B [ j^{A} , \D_A  j^C ] -   \D_B j_C [ j^{A}, \D_A  j^{B}]   )\\
& = 
- \tr( [ j^{A }  , j^{B}   ] \D_A \D_B  j^C  ) =  -\tfrac{1}{2}  \Fddu{A}{B}{D} \tr( [ j^{A }  , j^{B}   ] \D_D j^C  ) = 0,
}{dKkin_zero}
where the last equality holds by the identity in \Eq{FKKD}. Hence, the kinematic current is conserved on the support of the equations of motion.  As we will see shortly, it is conservation of the kinematic current that actually enforces the kinematic Jacobi identities in the NLSM.

Finally, let us remark on the fact that color-kinematics duality of the NLSM in curved spacetime hinges so critically on the assumption that the spacetime manifold is symmetric.  This is particularly amusing because color-kinematics duality of the NLSM also requires that the internal field space manifold is symmetric.  Furthermore, this same condition is needed for the Adler zero \cite{Cheung:2020tqz,Cheung:2021yog}.  We leave for future work whether there is any deep significance carried by these mirrored requirements on the spacetime manifold and the internal manifold.

\subsubsection{Kinematic Jacobi Identities}

For the reasons we have just described, color-kinematics duality is automatically exhibited by the numerators in the four-point on-shell correlator of the NLSM in \Eq{NLSM_amps}.  As discussed more generally in \Eq{num_denom}, it is always possible to commute all propagators to the left in a local theory.  Implementing this in \Eq{NLSM_amps}, we obtain a well-defined split between numerators and denominators despite the fact that the isometric momenta are non-commutative.  The resulting form of the four-point on-shell correlator turns out to be color-kinematic dual,
\eq{
\AmpD_4 =  \frac{c_s}{\D_{12}^2} \mathbb{n}_s + \frac{c_t}{\D_{23}^2} \mathbb{n}_t + \frac{c_u}{\D_{31}^2} \mathbb{n}_u,
}
where the color structures are
\eq{
c_s = f_{a_1 a_2 b} \fudd{b}{ a_3 }{a_4}, \quad c_t = f_{ a_2 a_3 b} \fudd{b}{ a_1 }{a_4}, \quad c_u = f_{a_3 a_1 b} \fudd{b}{ a_2 }{a_4},
}
and the kinematic numerators are 
\eq{
\mathbbm{n}_{s} &=\frac{1}{q \cdot \D_4} \left[ (\D_3 \cdot  \D_{12})  (\D_1\cdot \D_2) (q\cdot \D_1- q\cdot  \D_2) - (\D_1\cdot \D_2) (\D_1\cdot \D_3- \D_2\cdot \D_3)   (q\cdot \D_3) \right]
\\
\mathbbm{n}_t &= \frac{1}{q \cdot \D_4} \left[(\D_1\cdot  \D_{23})  (\D_2\cdot \D_3) (q\cdot \D_2- q\cdot \D_3) - (\D_2\cdot \D_3) (\D_2 \cdot \D_1- \D_3\cdot \D_1)  (q\cdot \D_1) \right]
\\
\mathbbm{n}_u &=\frac{1}{q \cdot \D_4} \left[( \D_2\cdot  \D_{31})  (\D_3\cdot \D_1) (q\cdot \D_3- q\cdot \D_1) - (\D_3\cdot \D_1) (\D_3 \cdot \D_2- \D_1\cdot \D_2)   (q\cdot \D_2) \right]  .
}{kin_num_4}
As advertised, these numerators satisfy the Jacobi relation
\eq{
\mathbbm{n}_{s} + \mathbbm{n}_{t}+\mathbbm{n}_{u} 
=& \frac{1}{q \cdot \D_4}  F_{ABC} \left[ \D_3^A \D_2^B  \D^C_1 (q\cdot \D_1) + \D_1^A \D_3^B\D^C_2 (q\cdot \D_2) + \D_2^A \D_1^B  \D^C_3 (q\cdot \D_3)\right]
 = 0.
}
In the special case of AdS in \cite{DiwakarHRT21}, it was shown explicitly that all terms involving a single factor of $F_{ABC}$ identically vanish, thus yielding zero.  This is actually true for symmetric spacetimes as well.  To understand why the right-hand side of the above equation is zero, let us consider the more general case of off-shell color-kinematics duality in the NLSM.

Specifically, we consider an internal off-shell four-point subdiagram computed using our NLSM formulation.
This subdiagram is essentially identical to the four-point on-shell correlator computed previously except that the external legs 1,2,3,4 are replaced with arbitrary sets of isometric momenta, $I,J,K,L$, and the external chiral currents are left uncontracted.    This subdiagram receives contributions in three channels,
\eq{
        &
\raisebox{-42pt}{
\begin{tikzpicture}[scale=0.5]
    \draw [internalleg] (3.75,2.1) -- (2.625,+1.05) ;
    \draw [internalleg] (3.75,-2.1)-- (1.5,0) ;
    \draw [internalleg] (2.625,+1.05) --  (1.5,0) ;
    \draw [internalleg] (3.75,0)   -- (2.625,+1.05) ;
    \draw [internalleg] (1.5,0) -- (0,0)   ;
    \node[left] at (0,0) {\small $L$};
    \node[right] at (3.75,2.1) {\small $I$};
    \node[right] at (3.75,0) {\small $J$};
    \node[right] at (3.75,-2.1) {\small $K$};
\end{tikzpicture} 
} + 
\raisebox{-42pt}{
\begin{tikzpicture}[scale=0.5]
    \draw [internalleg] (3.75,2.1) -- (1.5,0) ;
    \draw [internalleg] (3.75,-2.1)-- (2.625,-1.05) ;
    \draw [internalleg] (2.625,-1.05) --  (1.5,0) ;
    \draw [internalleg] (3.75,0)   -- (2.625,-1.05) ;
    \draw [internalleg] (1.5,0) -- (0,0)   ;
    \node[left] at (0,0) {\small $L$};
    \node[right] at (3.75,2.1) {\small $I$};
    \node[right] at (3.75,0) {\small $J$};
    \node[right] at (3.75,-2.1) {\small $K$};
\end{tikzpicture} 
}
+ 
\raisebox{-42pt}{
\begin{tikzpicture}[scale=0.5]
    \draw [internalleg] (3.75,2.1) -- (2.625,+1.05) ;
    \draw [internalleg] (3.75,-2.1)-- (1.5,0) ;
    \draw [internalleg] (2.625,+1.05) --  (1.5,0) ;
    \draw [internalleg] (3.75,0)   -- (2.625,+1.05) ;
    \draw [internalleg] (1.5,0) -- (0,0)   ;
    \node[left] at (0,0) {\small $L$};
    \node[right] at (3.75,2.1) {\small $I$};
    \node[right] at (3.75,0) {\small $K$};
    \node[right] at (3.75,-2.1) {\small $J$};
\end{tikzpicture} }
       \\ \\ & \quad =  \frac{c_s}{\D_{IJ}^2} \mathbb{n}_s^{A_I A_J A_K A_L} + \frac{c_t}{\D_{JK}^2} \mathbb{n}_t^{A_J A_K A_I A_L} + \frac{c_u}{\D_{KI}^2} \mathbb{n}_u^{A_K A_I A_J A_L},
}
where the tensor kinematic numerators are
\eq{
\mathbb{n}_s^{A_I A_J A_K A_L} &= \D_{IJ}^{A_K} \D_I^{A_J} g^{A_I A_L} + \D_J^{A_I} \D_K^{A_J} g^{A_K A_L} - \{ I\leftrightarrow J \} \\
\mathbb{n}_t^{A_I A_J A_K A_L} &= \D_{JK}^{A_I} \D_J^{A_K} g^{A_J A_L} + \D_K^{A_J} \D_I^{A_K} g^{A_I A_L} - \{ J\leftrightarrow K \} \\
\mathbb{n}_u^{A_I A_J A_K A_L} &= \D_{KI}^{A_J} \D_K^{A_I} g^{A_K A_L} + \D_I^{A_K} \D_J^{A_I} g^{A_J A_L} - \{ K\leftrightarrow I \} ,
}
which satisfy the kinematic Jacobi identities off-shell,
\eq{
&\mathbb{n}_s^{A_I A_J A_K A_L} + \mathbb{n}_t^{A_I A_J A_K A_L}+ \mathbb{n}_u^{A_I A_J A_K A_L}= \D_{I}^{[A_J} \D_I^{A_K]} g^{A_I A_L}+ \D_{J}^{[A_K} \D_J^{A_I]} g^{A_J A_L} +\D_{K}^{[A_I} \D_K^{A_J]} g^{A_K A_L}\\
&= \Fuud{A_J}{A_K}{B} \D^B_I g^{A_I A_L} + \Fuud{A_J}{A_K}{B} \D^B_I g^{A_I A_L} + \Fuud{A_J}{A_K}{B} \D^B_I g^{A_I A_L} = 0.
}{kin_jac_gen}
It is clear that the numerators of the four-point on-shell correlator in \Eq{kin_num_4} are obtained by contracting the above expression with the appropriate external polarizations.

In  \Eq{kin_jac_gen}, the sum of terms linear in the structure constants actually vanishes.  The reason for this is that these terms are exactly equal to the Feynman rule corresponding to the right-hand side of the kinematic current conservation equation in \Eq{dKkin_zero}.  Hence, once the free indices in \Eq{kin_jac_gen}  are contracted against other tensors to yield an isometric invariant, the resulting expression will be zero.
As described in flat space \cite{CCK}, this occurs because kinematic current conservation exactly mandates the kinematic Jacobi identity.

\subsection{Special Galileon}

The above analysis suggests a clear-cut strategy for constructing the double copy of the NLSM in curved spacetime.  Based on what is known in flat space \cite{Cheung:2020djz,CCK}, the resulting theory should be the special Galileon (SG) \cite{Cheung:2014dqa, Hinterbichler:2015pqa}, appropriately generalized to curved spacetime \cite{BonifacioHJR18,BonifacioHJR21}.  To this end, we define a generalization of the chiral tensor of the SG \cite{CCK} on a symmetric manifold,
\eq{
j^{A\ov A} = j^{\mu\ov \mu}  K_\mu^A K_{\ov \mu}^{\ov A}.
}
Applying the replacement in \Eq{sub_struct} to the color structure constants in the equation of motion for the NLSM in \Eq{EOM_NLSM_iso},  we obtain a conjectural equation of motion for the SG,
\eq{
\D^2 j^{C\ov C} +j^{A \ov A} \D_A \D_{\ov A} j^{C \ov C} -\D_A j^{C \ov A} \D_{\ov A} j^{A\ov C}=0.
}{SG_conjecture}
Naively, ordering of the Lie derivatives with barred and unbarred isometric indices is ambiguous in \Eq{SG_conjecture}. However, this ambiguity actually disappears because 
\eq{
j^{A \ov A} \D_{[A} \D_{\ov A]} j^{C \ov C} = j^{A \ov A} \Fddu{A}{\ov A}{B} \D_B j^{C \ov C} =0,
}
which vanishes due to the identity in \Eq{FKKD} for symmetric spacetimes.

The natural next step is to then compute the on-shell correlators derived from \Eq{SG_conjecture}.  Since \Eq{SG_conjecture} was obtained through a direct replacement of color with kinematics in the NLSM, the corresponding SG propagator, interaction vertex, and external polarizations are literally the square of those of the NLSM in \Eq{NLSM_Feynman_Rules}.  In particular, this implies that the leaf leg polarizations should be $(\D_I)^{A_I}(\D_I)^{\ov A_I}$, while the root leg polarizations should be $q_{A_I} q_{{\ov A}_I} / ( q\cdot \D)^2$.

Unfortunately, in implementing these Feynman rules we immediately encounter a confusion.  In particular, recall that in the color-kinematic dual formulation of the NLSM described in \Eq{EOM1st_NLSM_iso_comp}, conservation of the chiral current, $\D_A j^{aA} =0$, is critical.   Implicitly, this constraint ensured that the reference vector $q$, appearing in the root leg polarization in \Eq{NLSM_Feynman_Rules} and in the map from the chiral current to the NLSM scalar in \Eq{pi_from_j}, is arbitrary and does not affect the final answer.  By taking the divergence of the NLSM equation of motion in \Eq{EOM_NLSM_iso}, one  can verify explicitly that $\D_A j^{aA} =0$.

Analogously, it is similarly necessary that all $q$ dependence also drops out from the final answer in the SG.   Thus one expects that the chiral tensor should be conserved on both indices $\D_A j^{A \ov {A}} = \D_{\ov A}j^{A\ov A} = 0$.   Taking the divergence of the conjectured SG equation of motion in \Eq{SG_conjecture}, we obtain 
\eq{
\D_C\left( \D^2 j^{C\ov C} +j^{A \ov A} \D_A \D_{\ov A} j^{C \ov C} -\D_A j^{C \ov A} \D_{\ov A} j^{A\ov C}\right)
=
F_{C D E}\Fudd{E}{\ov B}{ B}\D^B \left( j^{C \ov C}j^{D \ov B} \right)-\frac{1}{2} \D^B\left(j^{\ov B \ov C} j_{B \ov B} \right).
}
The above expression is not automatically zero, and so without imposing this additional constraint by fiat, \Eq{SG_conjecture} appears inconsistent with conservation of the chiral tensor. We leave exploration of this apparent incompatibility and our conjectural SG formulation to further work.

\subsection{Fundamental BCJ Relation}

We are now equipped to derive the fundamental BCJ relation for BAS theory and the NLSM in curved spacetime.   As we will see, the  fundamental BCJ relation \cite{Bern:2008qj} is physically equivalent to conservation of the kinematic current.   Our approach follows closely the argument outlined in \cite{CCK} for flat space on-shell scattering amplitudes.

\subsubsection{Null Color Replacement}

To begin, recall from our discussion of BAS theory that the dual color current is conserved on account of the corresponding Jacobi identities. In components, this conservation equation is
\eq{
0=\D^\mu {\cal K}^{\ov c}_\mu =  \fbarddu{ a}{ b}{ c}  \tr(\phi^{ \ov a} \lrsq{\D} \phi^{ \ov b} ) =
\fbarddu{ a}{ b}{ c}  \delta_{ab}  \phi^{ a\ov a}  \lrsq{\D} \phi^{b \ov b} ,
}{dK_comp}
which is exceedingly similar in structure to the interaction term in the BAS theory equation of motion in \Eq{EOM_BAS_iso},
\eq{
\fbarddu{ a}{ b}{ c}  [\phi^{\ov a} , \phi^{\ov b} ] = \fbarddu{ a}{ b}{ c}   \fddu{ a}{ b}{ c}  \phi^{a\ov a} \phi^{b\ov b} .
}
In fact, these expressions are literally identical up to a replacement,
\eq{
\fddu{ a}{ b}{ c} \quad  \rightarrow \quad \delta \fddu{ a}{ b}{ c} =   \delta_{ab}   \lrsq{\D},
}{color_shift}
where we dub $\delta \fddu{ a}{ b}{ c}$ the {\it null color structure constant} for reasons that will become apparent shortly.
Here the left and right arrows on the Killing Casimir indicate whether it acts on the field with the color index $a$ or $b$, respectively.\footnote{The null color structure constant defined in \Eq{color_shift} has no dependence on the color index $c$.  While this is peculiar, it will not matter because we will never actually contract $c$ with other indices.  In any case, one can alternatively define the null color structure constant to be $ \delta \fddu{ a}{ b}{ c} =   q^c \delta_{ab}   \lrsq{\D}$, where $q^c$ is an arbitrary color reference.   With this definition, all of our subsequent analyses will still apply. }  We will refer to \Eq{color_shift} and similar manipulations as a {\it null color replacement}.
Notice that the null color structure constant, $\delta \fddu{ a}{ b}{ c}$, is antisymmetric on its first and second indices, as would be expected for the structure constant of an actual underlying algebra.  The null color replacement in \Eq{color_shift} is the curved spacetime generalization of the color factor transformation \cite{Bern:2008qj,Brown:2016hck} discovered in flat space amplitudes, and later used in the context of equations of motion to prove the fundamental BCJ relations \cite{CCK}.

The above observation gives a simple perturbative method for explicitly computing the divergence of the dual color current---which should be zero. In particular, we evaluate $\D^\mu {\cal K}^{\ov c}_\mu$ by computing the right-hand side of \Eq{dK_comp} using Berends-Giele recursion.   However, as we have just shown, the right-hand side of \Eq{dK_comp} is literally the same as the usual interaction vertex of BAS theory up to the null color replacement in \Eq{color_shift}.   This means that $\D^\mu {\cal K}^{\ov c}_\mu$ can be obtained by computing an on-shell correlator of BAS theory and then applying  \Eq{color_shift} to the color structure constant connected to the root leg.  Here the root leg makes an appearance because it labels the location of the insertion of the operator being evaluated, namely $\D^\mu {\cal K}^{\ov c}_\mu$.  After applying the null color replacement, the resulting expression will thus vanish.

To implement the procedure described above, we take the following mechanical steps.    First, compute the $n$-point on-shell correlator $A_n$ of BAS theory.  Second, apply the null color replacement to the $n$-th, {\it i.e.}~the root leg,
\eq{
\fddu{a_I}{a_J}{a_n}  \quad \rightarrow \quad \delta \fddu{ a_I}{ a_J}{ a_n} =\delta_{a_I a_J}(\D^2_J -\D^2_I). 
}{f_shift}
Here $a_n$ is the color index of the root leg while  $a_I$ and $a_J$ are the color indices of the pair of fields that interact directly with the root leg at a cubic vertex.  The kinematic factors $D^2_I$ and $D^2_J$ denote the squares of the total isometric momenta flowing into each of those fields.  Our claim is then that the resulting expression vanishes
\eq{
A_n \quad\rightarrow\quad \delta A_n = 0,
}
and similarly for the corresponding isometric representations,
\eq{
\AmpD_n \quad\rightarrow\quad \delta \AmpD_n = 0.
}
These expressions are all proportional to the divergence of the dual color current, $\D^\mu {\cal K}^{\ov c}_\mu$, and thus vanish on the support of the equations of motion. 

Rather incredibly, the null color structure constants still satisfy the Jacobi identity when contracted with regular color structure constants.  In particular,
\eq{
 \fddu{a_I}{a_J}{b}  \delta  \fddu{b}{a_K}{a_n}    +\fddu{a_J}{a_K}{b}  \delta  \fddu{b}{a_I}{a_n}  +\fddu{a_K}{a_I}{b}  \delta  \fddu{b}{a_J}{a_n} &= f_{a_I a_J a_K} \left( \D^2_{IJ} -\D^2_K+\D^2_{JK} -\D^2_I+ \D^2_{KI} -\D^2_J\right)\\
 &= f_{a_I a_J a_K} \D^2_{IJK} = f_{a_I a_J a_K} \D^2_{n}=0,
 }
 where all we have used is total isometric momentum conservation and the isometric on-shell condition for the $n$-th leg.  This implies that the substitution in \Eq{color_shift} commutes the Jacobi identities.  In other words, we can reshuffle color structures in the on-shell correlators of BAS theory with impunity, and then afterwards apply \Eq{color_shift} and the resulting expression will still vanish.  As we will see shortly, this directly implies the fundamental BCJ relation.  
 
Conveniently, the above logic trivially generalizes from BAS theory to the NLSM.    This is clear because the kinematic current of the NLSM is also conserved,
\eq{
0 = \D^A {\cal K}_{A}^C & = \delta_{ab}   j^{a B } \lrsq{\D}  \D_B j^{bC  },
}{DK_NLSM_zero}
where the right-hand side can also be obtained by applying the null color replacement in \Eq{color_shift} to the interaction term of the NLSM equation of motion,
\eq{
{} [ j^{A},  \D_A  j^{C}] &= \fddu{a}{b}{c} j^{a B }  \D_B j^{bC  }.
}
There is no ambiguity in the ordering of Lie derivatives in \Eq{DK_NLSM_zero}, since the Killing Casimir $\D^2$ commutes with everything.  As a result, the claim made above---that the null color replacement of any $n$-point on-shell correlator equals zero---applies equally well for the NLSM as well as BAS theory.

 In what follows, we study some simple explicit examples from BAS theory at four- and five-point and then present a general derivation of the fundamental BCJ relation at $n$-point.  Our proof is sufficiently general that it applies to both BAS theory as well as the NLSM.

\subsubsection{Explicit Examples}

Starting from the four-point on-shell correlator of BAS theory in \Eq{BAS_Amp}, we apply the null color replacement in \Eq{color_shift} to the root leg to obtain 
\eq{
\delta \AmpD_4  &=    f_{a_1 a_2 a_3}  \fbarddu{a_1}{a_2}{b} \fbarddu{b}{a_3}{a_4} \frac{1}{\D^2_{12}}( \D^2_{12} - \D^2_3 )+ \textrm{2 terms}
 \\
&=   f_{a_1 a_2 a_3} (\fbarddu{a_1}{a_2}{b} \fbarddu{b}{a_3}{a_4} +\fbarddu{a_2}{a_3}{b} \fbarddu{b}{a_1}{a_4}  +\fbarddu{a_3}{a_1}{b}  \fbarddu{b}{a_2}{a_4})   =0.
}{dG4_AdS}
Rearranging terms on the right-hand side, we can recast the vanishing of $\delta \AmpD$ as a relation among the color-stripped correlators
\eq{
\delta \AmpD_4  
= & \phantom{{}+{}} f_{a_1 a_2 a_3} ( \D^2_{12} - \D^2_3 )\left[ \fbarddu{a_1}{a_2}{b} \fbarddu{b}{a_3}{a_4}  \frac{1}{\D_{12}^2} - \fbarddu{a_3}{a_1}{b}  \fbarddu{b}{a_2}{a_4}  \frac{1}{\D_{31}^2}  \right]\\
&+f_{a_3 a_1 a_2} ( \D^2_{23 } - \D^2_1 )\left[\fbarddu{a_2}{a_3}{b} \fbarddu{b}{a_1}{a_4} \frac{1}{\D^2_{23}}- \fbarddu{a_3}{a_1}{b}  \fbarddu{b}{a_2}{a_4}   \frac{1}{\D^2_{31}} \right]\\
= &  \phantom{{}+{}} f_{a_1 a_2a_3}  \left[ \D_{23}^2  \AmpD(1423) -\D^2_{12} \AmpD(1243)  \right],
}{dG4_BCJ}
which is the fundamental BCJ relation for the four-point on-shell correlator.

Applying the same logic to the five-point on-shell correlator for BAS theory in \Eq{BAS_Amp}, we apply the null color replacement in \Eq{color_shift} to obtain 
 \eq{
\delta \AmpD_5  =&   \phantom{{}+{}}    \fddu{a_1}{a_2}{b}  f_{b a_3 a_4}    \fbarddu{a_1}{a_2}{b_1}\fbarddu{b_1}{a_3}{b_2}  \fbarddu{b_2}{a_4}{a_5}   (\D^2_{123} -\D_4^2) \frac{1}{\D_{123}^2 \D^2_{12}}  + \textrm{11 terms} \\
&+    \fddu{a_1}{a_2}{b}  f_{a_3 a_4 b}   \fbarddu{a_1}{a_2}{b_1}  \fbarddu{b_2}{b_1}{a_5}\fbarddu{a_3}{a_4}{b_2}    (\D^2_{34} - \D^2_{12})\frac{1}{\D_{12}^2 \D_{34}^2} + \textrm{2 terms} \\
=&  \phantom{{}+{}}\fddu{a_1}{a_2}{b}  f_{a_3 a_4 b}  \left[ \D_{234}^2 \AmpD(15234) +(\D_{34}^2-\D_{12}^2) \AmpD(12534)  - \D_{123}^2 \AmpD(12354) \right] +\cdots,
 }{dG5_AdS}
where in the last line we have reduced to the four-point DDM basis \cite{DDM} for color and the ellipses denote other independent color structures.  If we demand that the right-hand side is zero, then the coefficient of each independent color structure factor must vanish.  The resulting identity is precisely the fundamental BCJ relation for the five-point on-shell correlator.

\subsubsection{Proof of the Fundamental BCJ Relation}

We are now ready to prove the fundamental BCJ relation in general.
To begin, let us decompose the color structures of the $n$-point on-shell correlator in BAS theory or the NLSM into the DDM basis, yielding
\eq{
\AmpD_n = \sum_{\sigma \in S_{n-2}} f(1, \sigma, n-1) \AmpD(1, \sigma, n-1).
}
Here the color-stripped correlators on the right-hand side encode all implicit dependence on kinematics or dual color structures.  Meanwhile, the half-ladder color structures are
\eq{
&f(1,2,
\cdots, i, n, i+1, \cdots , n-1) =   \raisebox{-4pt}{
\begin{tikzpicture}
            \draw [very thick](-2.0,0) -- (+2.0,0);
            \draw [very thick](-1.5,0) -- (-1.5,0.5);
            \draw [very thick](-0.5,0) -- (-0.5,0.5);
            \draw [very thick](-0.0,0) -- (-0.0,0.5);
            \draw [very thick](+0.5,0) -- (+0.5,0.5);
            \draw [very thick](+1.5,0) -- (+1.5,0.5);
            \node[left]  at (-2.0,0.0) {\scriptsize$1$};
            \node[above] at (-1.5,0.5) {\scriptsize$2$};
            \node[above] at (-0.5,0.5) {\scriptsize$i$};
            \node[above] at (-0.0,0.5) {\scriptsize$n$};
            \node[above] at (+0.5,0.5) {\scriptsize $i+1$};
            \node[above] at (+1.5,0.5) {\scriptsize $n-2$};
            \node[right] at (+2.0,0.0) {\scriptsize$n-1$};
            \node at (-1.0,0.3) {\scriptsize$\cdots$};
            \node at (+1.0,0.3) {\scriptsize$\cdots$};
\end{tikzpicture}} \\
&
\hspace{1cm}=\fddu{a_1}{a_2}{b_1} 
\fddu{b_1}{a_3}{b_2}  
\cdots  \fddu{b_{i-2}}{ a_{i}}{ b_{i-1}}\fddu{b_{i}}{b_{i-1} }{  a_n}  \fddu{a_{i+1}}{ b_{i+1}}{  b_{i}}  
\cdots 
\fddu{a_{n-3}}{b_{n-3}}{ b_{n-4} }
\fddu{a_{n-2}}{a_{n-1}}{b_{n-3} }.   
}
We have purposely chosen a DDM decomposition in which legs 1 and $n-1$ are at the endcaps, so the root leg $n$ always appears somewhere in the bulk of each half-ladder.

Under the null color replacement in \Eq{color_shift} to the root leg, the half-ladder maps to 
\eq
{
\delta f(1,2,3,\cdots, i, n, i+1, \cdots , n-1)&=   f(1,2,3,\cdots, i, i+1, \cdots , n-1) \,( \D^2_{1\cdots i} - \D^2_{i+1 \cdots n-1} )  .
}{delta_ladder}
Note that we started with $(n-2)!$ independent half-ladders in the original $n$-point correlator, but the resulting expression involves half-ladders associated with the $(n-1)$-point correlator, of which $(n-3)!$ are independent. Crucially, legs 1 and $n-1$ are at the endcaps of every half-ladder, so the right-hand side of \Eq{delta_ladder} is already written in a basis of independent color structures.  

As argued previously, applying the null color shift to the $n$-point on-shell correlator should yield zero due to conservation of the dual color current, so
\eq{
\delta \AmpD_n =  f(1,2,3,\cdots, n-1) \sum_{i=1}^{n-2} ( \D^2_{i+1 \cdots n-1}- \D^2_{1\cdots i}  ) \AmpD(1,\cdots, i, n, i+1, \cdots , n-1) +\cdots =0.
}{BCJ_flat2}
The terms in ellipses are other independent half-ladder color structures.  
The coefficient of each color structure must vanish, yielding
\eq{
 \sum_{i=1}^{n-2} (\D^2_{i+1 \cdots n-1}- \D^2_{1\cdots i}  ) \AmpD(1,\cdots, i, n, i+1, \cdots , n-1) =0,
}
which is exactly the fundamental BCJ relation in curved spacetime.  This result applies for both BAS theory and the NLSM on any symmetric manifold.  The above argument mirrors exactly the proof of the fundamental BCJ relation in flat space from \cite{CCK}, and proves the conjecture in AdS proposed in \cite{DiwakarHRT21}.

\section{Conclusions}

In this work we have generalized flat space on-shell kinematics and on-shell scattering amplitudes to curved symmetric spacetimes.   Remarkably, the resulting kinematics are essentially identical to flat space except that the corresponding momenta---which are isometry-generating Lie derivatives---are non-commutative.  In the process we  have also shown that on-shell correlators are essentially identical to their flat space cousins, albeit with a canonical ordering of momenta fixed by locality.   Furthermore, we have shown that on-shell correlators are invariant under changes of field basis.   Finally, we have explicitly computed on-shell correlators in BAS theory and the NLSM, constructed the corresponding color-kinematics duality at the level of fields, and proven the fundamental BCJ relation.

Our results leave open a number of avenues for future inquiry.  First and foremost is the question of generalization to include spin.   It seems quite likely that it is possible to transform theories with spin into isometric frame and construct a corresponding formulation of on-shell kinematics which includes external polarizations. 

It is also natural to ask whether the isometric representation for on-shell correlators is easily extended to include loops.  Obviously, much of our discussion has leaned heavily on equations of motion for computing the classical sector of the dynamics.  It would be interesting to see if this can be extended, perhaps in line with recent work in AdS \cite{Herderschee21} and dS \cite{Gomez:2021ujt}.  A closely related question is whether the machinery we have presented can be used to compute off-shell correlators, corresponding to processes with local insertions of a source. 

Given that isometric kinematics are an efficient formalism for computing, it would be interesting to apply these tools to obtain higher-point data for boundary correlators in AdS.  While we have implicitly focused on Euclidean signature, our results may extend to Lorentzian AdS via analytic continuation.    Results of this kind might in turn be useful for studying momentum space, Mellin space, and harmonic analysis methods at higher points \cite{FitzpatrickKPRV11,Paulos11,Parikh19,MeltzerPS19,JepsenP19,AlbayrakCK20,Parikh19B}. It would also be interesting to use isometric kinematics to compute diagrams via bootstrap methods rather than by direct computation. Comparison to existing methods of obtaining correlators from weight shifting operators may be illuminating as well \cite{KarateevKS17,CostaH18}.

Since our results apply to large classes of symmetric spacetimes, it is natural to consider their implications for dS space.  For this case we expect that our on-shell correlators correspond to coefficients of the late-time wavefunction of the universe. For recent work on analytic continuations from AdS to dS, see \cite{SleightT19,SleightT21,SleightT20,Meltzer21,HogervorstPV21}. Our results may be useful for bootstrapping dS quantities in concert with the methods in \cite{Arkani-HamedBLP18,BaumannDJLP19,BaumannDJLP20,BaumannCDJLP21,DiPietro21}. Isometric kinematics may simplify aspects of momentum-space computations in dS. It would be interesting to explore how our results connect to recent work on weight-shifting operators and differential representations \cite{BaumannDJLP19,HillmanP21}. 

It is also interesting to ask whether our results can be understood using the WZW-like ambitwistor strings of Ref.~\cite{Roehrig:2020kck} of scattering equation approach of Ref.~ \cite{EberhardtKM20}, extended beyond AdS to more general symmetric spaces. 

A final direction for future exploration is to extend our understanding of color-kinematics duality in curved spacetime to a broader class of theories.  We have derived the mapping between BAS theory and the NLSM, but there is still the question of how to construct the SG.  Perhaps even more pressing is the question of color-kinematics duality in gauge theory and gravity.

\begin{center} 
   {\bf Acknowledgments}
\end{center}
\noindent 
We are grateful to Andreas Helset, James Mangan, and David Meltzer for useful discussions and comments on the paper. 
We are especially indebted to Aidan Herderschee, Radu Roiban, and Fei Teng for comments on the paper and sharing their related preprint \cite{HRTpreprint}.
C.C.~and J.P.-M. are supported by the DOE under grant no.~DE-SC0011632 and by the Walter Burke Institute for Theoretical Physics.  A.S.~was supported by the NSF under grant no.~NSF-PHY/211673 and the College of
Arts and Sciences of the University of Kentucky, and would like to thank the Walter Burke Institute for
Theoretical Physics for hospitality during the completion of this project.

\appendix
\section{Embedding Space}
\label{sec:embed}

Let us implement the mapping to isometric frame using embedding space methods.\footnote{For a review of embedding space, see e.g. \cite{Penedones:2016voo}.}  Consider a flat $(d+2)$-dimensional embedding space with arbitrary signature, labelled by a coordinate $X^M$.  We define a $(d+1)$-dimensional surface in this embedding space via the constraint
\eq{
X_M X^M = X^2 = {\rm constant}.
}
For the case of a hyperboloid, the resulting manifold is $(d+1)$-dimensional AdS, which is dual to a $d$-dimensional CFT.

It will be useful to define a projector that zeros out components perpendicular to the surface,
\eq{
\Pi_{MN} = \eta_{MN} - \frac{X_M X_N }{X^2}.
}
The metric on the surface is given by the flat embedding space metric followed by a projection of all free indices onto the embedding surface,
\eq{
g_{MN} = \Pi \circ \eta_{MN} \qandq g^{MN} = \Pi \circ \eta^{MN} .
}
Similarly, covariant derivatives on the surface are given by partial derivatives in the embedding space followed by a projection,
\eq{
\nabla_M = \Pi \circ \partial_M .
}
So for example, the constraint that defines the surface is itself covariantly constant, so
\eq{
\nabla_M(X^2) = \Pi \circ \partial_M(X^2) = \Pi \circ (2 X_M) =0.
}
Meanwhile, we know that the isometries of the surface are rotations and boosts,
\eq{
K_{MN} = K_{\phantom{R} MN}^R \partial_R =  X_{[M} \partial_{N]}.
}
From this we can reverse engineer the components of the Killing vectors,
\eq{
K_{\phantom{R} MN}^R =  X_{[M} \delta_{N]}^R.
}{K_embed}
Note that we have chosen to label the Killing vectors with a pair of antisymmetric embedding space indices.  Hence, the Killing vector $K_{MN}$ plays the role of $K_A$ described in the body of the paper.  As before in \Eq{def_DA}, we define the Lie derivative with respect to the Killing vector,
\eq{
\D_{MN} = {\cal L}_{K_{MN}}.
} 
Computing the commutator of Lie derivatives, we obtain the embedding space version of \Eq{Killing_algebra_extra},
\eq{
{} [ \D_{MN} ,\D_{RS} ]  = F_{MNRS}^{\phantom{MNRS} TU}  \D_{TU},
}
where the structure constant is given by
\eq{
 F_{MNRS}^{\phantom{MNRS} TU} =- 4 \eta_{MNI}^{\phantom{MNI} J}\eta_{RSJ}^{\phantom{RSJ} K}\eta_{\phantom{TU}K}^{TU\phantom{K} I} ,
}{F_embed}
in terms of the unit-normalized projection operator 
\eq{
\eta_{MNRS} = \frac{1}{2}(\eta_{MR}\eta_{NS}- \eta_{MS}\eta_{NR}).
}
The structure constant $ F_{MNRS}^{\phantom{MNRS} TU}$ plays the role of $\Fddu{A}{B}{C}$ referenced earlier in this paper.  In embedding space, the definition of the Killing metric in \Eq{def_gAB} becomes
\eq{
g_{MNRS} = -\frac{1}{8d X^2}  F_{MNIJ}^{\phantom{MNIJ} KL}  F_{RSKL}^{\phantom{RSKL} IJ} = \frac{1}{2X^2} \eta_{MNRS},
} 
where in the second equality we have introduced a convenient normalization to simplify later expressions.  This is permitted since the overall normalization of the Killing metric is arbitrary.   Here the Killing metric $g_{MNRS}$ plays the role of $g_{AB}$ described before. 

We are now equipped to verify various identities using these embedding space expressions.  To begin, we can check the Killing equation in \Eq{Killing_equation}, which in embedding space is
\eq{
\nabla_{(M} \K_{N)}^{\phantom{N} RS} & = \Pi \circ \left(\delta^{[R}_{(M} \delta^{S]}_{N)} \right) =0.
}
As in \Eq{g_from_K}, the spacetime metric can be expressed as an outer product of Killing vectors contracted with the Killing metric, which in embedding space is
\eq{
g_{RSTU} K^{\phantom{M} RS}_M K^{\phantom{N} TU}_N = \eta_{MN} -\frac{X_M X_N}{X^2} = g_{MN}.
}
Last but not least, plugging \Eq{K_embed} and \Eq{F_embed} into,
\eq{
F^{MNRSTU} \K_{\phantom{I} MN}^I \K_{\phantom{J} RS}^J K_{\phantom{K} TU}^K = 0,
}
we verify that \Eq{FKKK} is indeed satisfied.

\newpage

\bibliographystyle{ssg}
\bibliography{refs}

\begingroup\raggedright\begin{thebibliography}{100}

\bibitem{Bern:1994zx}
Z.~Bern, L.~J. Dixon, D.~C. Dunbar, and D.~A. Kosower, ``{One loop n point
  gauge theory amplitudes, unitarity and collinear limits},'' {\em Nucl. Phys.
  B} {\bf 425} (1994) 217--260,
  \href{http://xxx.lanl.gov/abs/hep-ph/9403226}{{\tt hep-ph/9403226}}.

\bibitem{Bern:1994cg}
Z.~Bern, L.~J. Dixon, D.~C. Dunbar, and D.~A. Kosower, ``{Fusing gauge theory
  tree amplitudes into loop amplitudes},'' {\em Nucl. Phys. B} {\bf 435} (1995)
  59--101, \href{http://xxx.lanl.gov/abs/hep-ph/9409265}{{\tt hep-ph/9409265}}.

\bibitem{Bern:1995db}
Z.~Bern and A.~G. Morgan, ``{Massive loop amplitudes from unitarity},'' {\em
  Nucl. Phys. B} {\bf 467} (1996) 479--509,
  \href{http://xxx.lanl.gov/abs/hep-ph/9511336}{{\tt hep-ph/9511336}}.

\bibitem{Bern:1997sc}
Z.~Bern, L.~J. Dixon, and D.~A. Kosower, ``{One loop amplitudes for e+ e- to
  four partons},'' {\em Nucl. Phys. B} {\bf 513} (1998) 3--86,
  \href{http://xxx.lanl.gov/abs/hep-ph/9708239}{{\tt hep-ph/9708239}}.

\bibitem{Britto:2004nc}
R.~Britto, F.~Cachazo, and B.~Feng, ``{Generalized unitarity and one-loop
  amplitudes in N=4 super-Yang-Mills},'' {\em Nucl. Phys. B} {\bf 725} (2005)
  275--305, \href{http://xxx.lanl.gov/abs/hep-th/0412103}{{\tt
  hep-th/0412103}}.

\bibitem{Berger:2008sj}
C.~F. Berger, Z.~Bern, L.~J. Dixon, F.~Febres~Cordero, D.~Forde, H.~Ita, D.~A.
  Kosower, and D.~Maitre, ``{An Automated Implementation of On-Shell Methods
  for One-Loop Amplitudes},'' {\em Phys. Rev. D} {\bf 78} (2008) 036003,
  \href{http://xxx.lanl.gov/abs/0803.4180}{{\tt 0803.4180}}.

\bibitem{Badger:2017jhb}
S.~Badger, C.~Br\o{}nnum-Hansen, H.~B. Hartanto, and T.~Peraro, ``{First look
  at two-loop five-gluon scattering in QCD},'' {\em Phys. Rev. Lett.} {\bf 120}
  (2018), no.~9 092001, \href{http://xxx.lanl.gov/abs/1712.02229}{{\tt
  1712.02229}}.

\bibitem{Abreu:2017hqn}
S.~Abreu, F.~Febres~Cordero, H.~Ita, B.~Page, and M.~Zeng, ``{Planar Two-Loop
  Five-Gluon Amplitudes from Numerical Unitarity},'' {\em Phys. Rev. D} {\bf
  97} (2018), no.~11 116014, \href{http://xxx.lanl.gov/abs/1712.03946}{{\tt
  1712.03946}}.

\bibitem{Alday:2007hr}
L.~F. Alday and J.~M. Maldacena, ``{Gluon scattering amplitudes at strong
  coupling},'' {\em JHEP} {\bf 06} (2007) 064,
  \href{http://xxx.lanl.gov/abs/0705.0303}{{\tt 0705.0303}}.

\bibitem{Bern:2007ct}
Z.~Bern, J.~J.~M. Carrasco, H.~Johansson, and D.~A. Kosower, ``{Maximally
  supersymmetric planar Yang-Mills amplitudes at five loops},'' {\em Phys. Rev.
  D} {\bf 76} (2007) 125020, \href{http://xxx.lanl.gov/abs/0705.1864}{{\tt
  0705.1864}}.

\bibitem{Bern:2012di}
Z.~Bern, J.~J. Carrasco, L.~J. Dixon, M.~R. Douglas, M.~von Hippel, and
  H.~Johansson, ``{D=5 maximally supersymmetric Yang-Mills theory diverges at
  six loops},'' {\em Phys. Rev. D} {\bf 87} (2013), no.~2 025018,
  \href{http://xxx.lanl.gov/abs/1210.7709}{{\tt 1210.7709}}.

\bibitem{Arkani-Hamed:2012zlh}
N.~Arkani-Hamed, J.~L. Bourjaily, F.~Cachazo, A.~B. Goncharov, A.~Postnikov,
  and J.~Trnka, {\em {Grassmannian Geometry of Scattering Amplitudes}}.
\newblock Cambridge University Press, 4, 2016.

\bibitem{Arkani-Hamed:2013jha}
N.~Arkani-Hamed and J.~Trnka, ``{The Amplituhedron},'' {\em JHEP} {\bf 10}
  (2014) 030, \href{http://xxx.lanl.gov/abs/1312.2007}{{\tt 1312.2007}}.

\bibitem{Carrasco:2021otn}
J.~J.~M. Carrasco, A.~Edison, and H.~Johansson, ``{Maximal Super-Yang-Mills at
  Six Loops via Novel Integrand Bootstrap},''
  \href{http://xxx.lanl.gov/abs/2112.05178}{{\tt 2112.05178}}.

\bibitem{Bern:2012uf}
Z.~Bern, J.~J.~M. Carrasco, L.~J. Dixon, H.~Johansson, and R.~Roiban,
  ``{Simplifying Multiloop Integrands and Ultraviolet Divergences of Gauge
  Theory and Gravity Amplitudes},'' {\em Phys. Rev. D} {\bf 85} (2012) 105014,
  \href{http://xxx.lanl.gov/abs/1201.5366}{{\tt 1201.5366}}.

\bibitem{Bern:2012cd}
Z.~Bern, S.~Davies, T.~Dennen, and Y.-t. Huang, ``{Absence of Three-Loop
  Four-Point Divergences in N=4 Supergravity},'' {\em Phys. Rev. Lett.} {\bf
  108} (2012) 201301, \href{http://xxx.lanl.gov/abs/1202.3423}{{\tt
  1202.3423}}.

\bibitem{Bern:2012gh}
Z.~Bern, S.~Davies, T.~Dennen, and Y.-t. Huang, ``{Ultraviolet Cancellations in
  Half-Maximal Supergravity as a Consequence of the Double-Copy Structure},''
  {\em Phys. Rev. D} {\bf 86} (2012) 105014,
  \href{http://xxx.lanl.gov/abs/1209.2472}{{\tt 1209.2472}}.

\bibitem{Bern:2013qca}
Z.~Bern, S.~Davies, and T.~Dennen, ``{The Ultraviolet Structure of Half-Maximal
  Supergravity with Matter Multiplets at Two and Three Loops},'' {\em Phys.
  Rev. D} {\bf 88} (2013) 065007, \href{http://xxx.lanl.gov/abs/1305.4876}{{\tt
  1305.4876}}.

\bibitem{Bern:2013uka}
Z.~Bern, S.~Davies, T.~Dennen, A.~V. Smirnov, and V.~A. Smirnov, ``{Ultraviolet
  Properties of N=4 Supergravity at Four Loops},'' {\em Phys. Rev. Lett.} {\bf
  111} (2013), no.~23 231302, \href{http://xxx.lanl.gov/abs/1309.2498}{{\tt
  1309.2498}}.

\bibitem{Bern:2014sna}
Z.~Bern, S.~Davies, and T.~Dennen, ``{Enhanced ultraviolet cancellations in
  $\mathcal N=5$ supergravity at four loops},'' {\em Phys. Rev. D} {\bf 90}
  (2014), no.~10 105011, \href{http://xxx.lanl.gov/abs/1409.3089}{{\tt
  1409.3089}}.

\bibitem{Bern:2014lha}
Z.~Bern, S.~Davies, and T.~Dennen, ``{The Ultraviolet Critical Dimension of
  Half-Maximal Supergravity at Three Loops},''
  \href{http://xxx.lanl.gov/abs/1412.2441}{{\tt 1412.2441}}.

\bibitem{Bern:2017lpv}
Z.~Bern, M.~Enciso, J.~Parra-Martinez, and M.~Zeng, ``{Manifesting enhanced
  cancellations in supergravity: integrands versus integrals},'' {\em JHEP}
  {\bf 05} (2017) 137, \href{http://xxx.lanl.gov/abs/1703.08927}{{\tt
  1703.08927}}.

\bibitem{Bern:2017ucb}
Z.~Bern, J.~J.~M. Carrasco, W.-M. Chen, H.~Johansson, R.~Roiban, and M.~Zeng,
  ``{Five-loop four-point integrand of $N=8$ supergravity as a generalized
  double copy},'' {\em Phys. Rev. D} {\bf 96} (2017), no.~12 126012,
  \href{http://xxx.lanl.gov/abs/1708.06807}{{\tt 1708.06807}}.

\bibitem{Bern:2018jmv}
Z.~Bern, J.~J. Carrasco, W.-M. Chen, A.~Edison, H.~Johansson,
  J.~Parra-Martinez, R.~Roiban, and M.~Zeng, ``{Ultraviolet Properties of
  $\mathcal N = 8$ Supergravity at Five Loops},'' {\em Phys. Rev. D} {\bf 98}
  (2018), no.~8 086021, \href{http://xxx.lanl.gov/abs/1804.09311}{{\tt
  1804.09311}}.

\bibitem{Herrmann:2018dja}
E.~Herrmann and J.~Trnka, ``{UV cancellations in gravity loop integrands},''
  {\em JHEP} {\bf 02} (2019) 084,
  \href{http://xxx.lanl.gov/abs/1808.10446}{{\tt 1808.10446}}.

\bibitem{Bourjaily:2018omh}
J.~L. Bourjaily, E.~Herrmann, and J.~Trnka, ``{Maximally supersymmetric
  amplitudes at infinite loop momentum},'' {\em Phys. Rev. D} {\bf 99} (2019),
  no.~6 066006, \href{http://xxx.lanl.gov/abs/1812.11185}{{\tt 1812.11185}}.

\bibitem{Edison:2019ovj}
A.~Edison, E.~Herrmann, J.~Parra-Martinez, and J.~Trnka, ``{Gravity loop
  integrands from the ultraviolet},'' {\em SciPost Phys.} {\bf 10} (2021),
  no.~1 016, \href{http://xxx.lanl.gov/abs/1909.02003}{{\tt 1909.02003}}.

\bibitem{Cheung:2018wkq}
C.~Cheung, I.~Z. Rothstein, and M.~P. Solon, ``{From Scattering Amplitudes to
  Classical Potentials in the Post-Minkowskian Expansion},'' {\em Phys. Rev.
  Lett.} {\bf 121} (2018), no.~25 251101,
  \href{http://xxx.lanl.gov/abs/1808.02489}{{\tt 1808.02489}}.

\bibitem{Kosower:2018adc}
D.~A. Kosower, B.~Maybee, and D.~O'Connell, ``{Amplitudes, Observables, and
  Classical Scattering},'' {\em JHEP} {\bf 02} (2019) 137,
  \href{http://xxx.lanl.gov/abs/1811.10950}{{\tt 1811.10950}}.

\bibitem{Bern:2019nnu}
Z.~Bern, C.~Cheung, R.~Roiban, C.-H. Shen, M.~P. Solon, and M.~Zeng,
  ``{Scattering Amplitudes and the Conservative Hamiltonian for Binary Systems
  at Third Post-Minkowskian Order},'' {\em Phys. Rev. Lett.} {\bf 122} (2019),
  no.~20 201603, \href{http://xxx.lanl.gov/abs/1901.04424}{{\tt 1901.04424}}.

\bibitem{Antonelli:2019ytb}
A.~Antonelli, A.~Buonanno, J.~Steinhoff, M.~van~de Meent, and J.~Vines,
  ``{Energetics of two-body Hamiltonians in post-Minkowskian gravity},'' {\em
  Phys. Rev. D} {\bf 99} (2019), no.~10 104004,
  \href{http://xxx.lanl.gov/abs/1901.07102}{{\tt 1901.07102}}.

\bibitem{Bern:2019crd}
Z.~Bern, C.~Cheung, R.~Roiban, C.-H. Shen, M.~P. Solon, and M.~Zeng, ``{Black
  Hole Binary Dynamics from the Double Copy and Effective Theory},'' {\em JHEP}
  {\bf 10} (2019) 206, \href{http://xxx.lanl.gov/abs/1908.01493}{{\tt
  1908.01493}}.

\bibitem{Bern:2021dqo}
Z.~Bern, J.~Parra-Martinez, R.~Roiban, M.~S. Ruf, C.-H. Shen, M.~P. Solon, and
  M.~Zeng, ``{Scattering Amplitudes and Conservative Binary Dynamics at ${\cal
  O}(G^4)$},'' {\em Phys. Rev. Lett.} {\bf 126} (2021), no.~17 171601,
  \href{http://xxx.lanl.gov/abs/2101.07254}{{\tt 2101.07254}}.

\bibitem{Herrmann:2021lqe}
E.~Herrmann, J.~Parra-Martinez, M.~S. Ruf, and M.~Zeng, ``{Gravitational
  Bremsstrahlung from Reverse Unitarity},'' {\em Phys. Rev. Lett.} {\bf 126}
  (2021), no.~20 201602, \href{http://xxx.lanl.gov/abs/2101.07255}{{\tt
  2101.07255}}.

\bibitem{Herrmann:2021tct}
E.~Herrmann, J.~Parra-Martinez, M.~S. Ruf, and M.~Zeng, ``{Radiative classical
  gravitational observables at $ \mathcal{O} $(G$^{3}$) from scattering
  amplitudes},'' {\em JHEP} {\bf 10} (2021) 148,
  \href{http://xxx.lanl.gov/abs/2104.03957}{{\tt 2104.03957}}.

\bibitem{Bern:2021yeh}
Z.~Bern, J.~Parra-Martinez, R.~Roiban, M.~S. Ruf, C.-H. Shen, M.~P. Solon, and
  M.~Zeng, ``{Scattering Amplitudes, the Tail Effect, and Conservative Binary
  Dynamics at ${\cal O}(G^4)$},''
  \href{http://xxx.lanl.gov/abs/2112.10750}{{\tt 2112.10750}}.

\bibitem{Bern:2008qj}
Z.~Bern, J.~J.~M. Carrasco, and H.~Johansson, ``{New Relations for Gauge-Theory
  Amplitudes},'' {\em Phys. Rev. D} {\bf 78} (2008) 085011,
  \href{http://xxx.lanl.gov/abs/0805.3993}{{\tt 0805.3993}}.

\bibitem{Kawai:1985xq}
H.~Kawai, D.~C. Lewellen, and S.~H.~H. Tye, ``{A Relation Between Tree
  Amplitudes of Closed and Open Strings},'' {\em Nucl. Phys. B} {\bf 269}
  (1986) 1--23.

\bibitem{Bern:2010ue}
Z.~Bern, J.~J.~M. Carrasco, and H.~Johansson, ``{Perturbative Quantum Gravity
  as a Double Copy of Gauge Theory},'' {\em Phys. Rev. Lett.} {\bf 105} (2010)
  061602, \href{http://xxx.lanl.gov/abs/1004.0476}{{\tt 1004.0476}}.

\bibitem{BCJReview}
Z.~Bern, J.~J. Carrasco, M.~Chiodaroli, H.~Johansson, and R.~Roiban, ``{The
  Duality Between Color and Kinematics and its Applications},''
  \href{http://xxx.lanl.gov/abs/1909.01358}{{\tt 1909.01358}}.

\bibitem{Penedones10}
J.~Penedones, ``{Writing CFT correlation functions as AdS scattering
  amplitudes},'' {\em JHEP} {\bf 03} (2011) 025,
  \href{http://xxx.lanl.gov/abs/1011.1485}{{\tt 1011.1485}}.

\bibitem{FitzpatrickKPRV11}
A.~L. Fitzpatrick, J.~Kaplan, J.~Penedones, S.~Raju, and B.~C. van Rees, ``{A
  Natural Language for AdS/CFT Correlators},'' {\em JHEP} {\bf 11} (2011) 095,
  \href{http://xxx.lanl.gov/abs/1107.1499}{{\tt 1107.1499}}.

\bibitem{RastelliZ16}
L.~Rastelli and X.~Zhou, ``{Mellin amplitudes for $AdS_5\times S^5$},'' {\em
  Phys. Rev. Lett.} {\bf 118} (2017), no.~9 091602,
  \href{http://xxx.lanl.gov/abs/1608.06624}{{\tt 1608.06624}}.

\bibitem{RastelliZ17}
L.~Rastelli and X.~Zhou, ``{How to Succeed at Holographic Correlators Without
  Really Trying},'' {\em JHEP} {\bf 04} (2018) 014,
  \href{http://xxx.lanl.gov/abs/1710.05923}{{\tt 1710.05923}}.

\bibitem{Sleight19}
C.~Sleight, ``{A Mellin Space Approach to Cosmological Correlators},'' {\em
  JHEP} {\bf 01} (2020) 090, \href{http://xxx.lanl.gov/abs/1906.12302}{{\tt
  1906.12302}}.

\bibitem{SleightT19}
C.~Sleight and M.~Taronna, ``{Bootstrapping Inflationary Correlators in Mellin
  Space},'' {\em JHEP} {\bf 02} (2020) 098,
  \href{http://xxx.lanl.gov/abs/1907.01143}{{\tt 1907.01143}}.

\bibitem{SleightT20}
C.~Sleight and M.~Taronna, ``{From AdS to dS Exchanges: Spectral
  Representation, Mellin Amplitudes and Crossing},''
  \href{http://xxx.lanl.gov/abs/2007.09993}{{\tt 2007.09993}}.

\bibitem{SleightT21}
C.~Sleight and M.~Taronna, ``{From dS to AdS and back},'' {\em JHEP} {\bf 12}
  (2021) 074, \href{http://xxx.lanl.gov/abs/2109.02725}{{\tt 2109.02725}}.

\bibitem{Giddings99}
S.~B. Giddings, ``{The Boundary S matrix and the AdS to CFT dictionary},'' {\em
  Phys. Rev. Lett.} {\bf 83} (1999) 2707--2710,
  \href{http://xxx.lanl.gov/abs/hep-th/9903048}{{\tt hep-th/9903048}}.

\bibitem{BalasubramanianGL99}
V.~Balasubramanian, S.~B. Giddings, and A.~E. Lawrence, ``{What do CFTs tell us
  about Anti-de Sitter space-times?},'' {\em JHEP} {\bf 03} (1999) 001,
  \href{http://xxx.lanl.gov/abs/hep-th/9902052}{{\tt hep-th/9902052}}.

\bibitem{Raju10}
S.~Raju, ``{BCFW for Witten Diagrams},'' {\em Phys. Rev. Lett.} {\bf 106}
  (2011) 091601, \href{http://xxx.lanl.gov/abs/1011.0780}{{\tt 1011.0780}}.

\bibitem{CostaGP14}
M.~S. Costa, V.~Gon\c{c}alves, and J.~a. Penedones, ``{Spinning AdS
  Propagators},'' {\em JHEP} {\bf 09} (2014) 064,
  \href{http://xxx.lanl.gov/abs/1404.5625}{{\tt 1404.5625}}.

\bibitem{LiuPRS18}
J.~Liu, E.~Perlmutter, V.~Rosenhaus, and D.~Simmons-Duffin, ``{$d$-dimensional
  SYK, AdS Loops, and $6j$ Symbols},'' {\em JHEP} {\bf 03} (2019) 052,
  \href{http://xxx.lanl.gov/abs/1808.00612}{{\tt 1808.00612}}.

\bibitem{GiombiST17}
S.~Giombi, C.~Sleight, and M.~Taronna, ``{Spinning AdS Loop Diagrams: Two Point
  Functions},'' {\em JHEP} {\bf 06} (2018) 030,
  \href{http://xxx.lanl.gov/abs/1708.08404}{{\tt 1708.08404}}.

\bibitem{DiPietro21}
L.~Di~Pietro, V.~Gorbenko, and S.~Komatsu, ``{Analyticity and Unitarity for
  Cosmological Correlators},'' \href{http://xxx.lanl.gov/abs/2108.01695}{{\tt
  2108.01695}}.

\bibitem{HogervorstPV21}
M.~Hogervorst, J.~a. Penedones, and K.~S. Vaziri, ``{Towards the
  non-perturbative cosmological bootstrap},''
  \href{http://xxx.lanl.gov/abs/2107.13871}{{\tt 2107.13871}}.

\bibitem{FitzpatrickK11}
A.~L. Fitzpatrick and J.~Kaplan, ``{Unitarity and the Holographic S-Matrix},''
  {\em JHEP} {\bf 10} (2012) 032, \href{http://xxx.lanl.gov/abs/1112.4845}{{\tt
  1112.4845}}.

\bibitem{AldayC17}
L.~F. Alday and S.~Caron-Huot, ``{Gravitational S-matrix from CFT dispersion
  relations},'' {\em JHEP} {\bf 12} (2018) 017,
  \href{http://xxx.lanl.gov/abs/1711.02031}{{\tt 1711.02031}}.

\bibitem{MeltzerPS19}
D.~Meltzer, E.~Perlmutter, and A.~Sivaramakrishnan, ``{Unitarity Methods in
  AdS/CFT},'' {\em JHEP} {\bf 03} (2020) 061,
  \href{http://xxx.lanl.gov/abs/1912.09521}{{\tt 1912.09521}}.

\bibitem{Ponomarev19}
D.~Ponomarev, ``{From bulk loops to boundary large-N expansion},'' {\em JHEP}
  {\bf 01} (2020) 154, \href{http://xxx.lanl.gov/abs/1908.03974}{{\tt
  1908.03974}}.

\bibitem{MeltzerS20}
D.~Meltzer and A.~Sivaramakrishnan, ``{CFT unitarity and the AdS Cutkosky
  rules},'' {\em JHEP} {\bf 11} (2020) 073,
  \href{http://xxx.lanl.gov/abs/2008.11730}{{\tt 2008.11730}}.

\bibitem{Meltzer21}
D.~Meltzer, ``{The inflationary wavefunction from analyticity and
  factorization},'' {\em JCAP} {\bf 12} (2021), no.~12 018,
  \href{http://xxx.lanl.gov/abs/2107.10266}{{\tt 2107.10266}}.

\bibitem{Meltzer:2021bmb}
D.~Meltzer, ``{Dispersion Formulas in QFTs, CFTs, and Holography},'' {\em JHEP}
  {\bf 05} (2021) 098, \href{http://xxx.lanl.gov/abs/2103.15839}{{\tt
  2103.15839}}.

\bibitem{BaumannCDJLP21}
D.~Baumann, W.-M. Chen, C.~Duaso~Pueyo, A.~Joyce, H.~Lee, and G.~L. Pimentel,
  ``{Linking the Singularities of Cosmological Correlators},''
  \href{http://xxx.lanl.gov/abs/2106.05294}{{\tt 2106.05294}}.

\bibitem{Goodhew:2020hob}
H.~Goodhew, S.~Jazayeri, and E.~Pajer, ``{The Cosmological Optical Theorem},''
  {\em JCAP} {\bf 04} (2021) 021,
  \href{http://xxx.lanl.gov/abs/2009.02898}{{\tt 2009.02898}}.

\bibitem{Jazayeri:2021fvk}
S.~Jazayeri, E.~Pajer, and D.~Stefanyszyn, ``{From locality and unitarity to
  cosmological correlators},'' {\em JHEP} {\bf 10} (2021) 065,
  \href{http://xxx.lanl.gov/abs/2103.08649}{{\tt 2103.08649}}.

\bibitem{Melville:2021lst}
S.~Melville and E.~Pajer, ``{Cosmological Cutting Rules},'' {\em JHEP} {\bf 05}
  (2021) 249, \href{http://xxx.lanl.gov/abs/2103.09832}{{\tt 2103.09832}}.

\bibitem{Goodhew:2021oqg}
H.~Goodhew, S.~Jazayeri, M.~H. Gordon~Lee, and E.~Pajer, ``{Cutting
  cosmological correlators},'' {\em JCAP} {\bf 08} (2021) 003,
  \href{http://xxx.lanl.gov/abs/2104.06587}{{\tt 2104.06587}}.

\bibitem{DiwakarHRT21}
P.~Diwakar, A.~Herderschee, R.~Roiban, and F.~Teng, ``{BCJ Amplitude Relations
  for Anti-de Sitter Boundary Correlators in Embedding Space},''
  \href{http://xxx.lanl.gov/abs/2106.10822}{{\tt 2106.10822}}.

\bibitem{EberhardtKM20}
L.~Eberhardt, S.~Komatsu, and S.~Mizera, ``{Scattering equations in AdS: scalar
  correlators in arbitrary dimensions},'' {\em JHEP} {\bf 11} (2020) 158,
  \href{http://xxx.lanl.gov/abs/2007.06574}{{\tt 2007.06574}}.

\bibitem{CCK}
C.~Cheung and J.~Mangan, ``{Covariant color-kinematics duality},'' {\em JHEP}
  {\bf 11} (2021) 069, \href{http://xxx.lanl.gov/abs/2108.02276}{{\tt
  2108.02276}}.

\bibitem{BzowskiMS17}
A.~Bzowski, P.~McFadden, and K.~Skenderis, ``{Renormalised 3-point functions of
  stress tensors and conserved currents in CFT},'' {\em JHEP} {\bf 11} (2018)
  153, \href{http://xxx.lanl.gov/abs/1711.09105}{{\tt 1711.09105}}.

\bibitem{FarrowLM18}
J.~A. Farrow, A.~E. Lipstein, and P.~McFadden, ``{Double copy structure of CFT
  correlators},'' {\em JHEP} {\bf 02} (2019) 130,
  \href{http://xxx.lanl.gov/abs/1812.11129}{{\tt 1812.11129}}.

\bibitem{LipsteinM19}
A.~E. Lipstein and P.~McFadden, ``{Double copy structure and the flat space
  limit of conformal correlators in even dimensions},'' {\em Phys. Rev. D} {\bf
  101} (2020), no.~12 125006, \href{http://xxx.lanl.gov/abs/1912.10046}{{\tt
  1912.10046}}.

\bibitem{Zhou21}
X.~Zhou, ``{Double Copy Relation for AdS},''
  \href{http://xxx.lanl.gov/abs/2106.07651}{{\tt 2106.07651}}.

\bibitem{AldayBFZ21}
L.~F. Alday, C.~Behan, P.~Ferrero, and X.~Zhou, ``{Gluon Scattering in AdS from
  CFT},'' {\em JHEP} {\bf 06} (2021) 020,
  \href{http://xxx.lanl.gov/abs/2103.15830}{{\tt 2103.15830}}.

\bibitem{AlbayrakKM20}
S.~Albayrak, S.~Kharel, and D.~Meltzer, ``{On duality of color and kinematics
  in (A)dS momentum space},'' {\em JHEP} {\bf 03} (2021) 249,
  \href{http://xxx.lanl.gov/abs/2012.10460}{{\tt 2012.10460}}.

\bibitem{Sivaramakrishnan21}
A.~Sivaramakrishnan, ``{Towards Color-Kinematics Duality in Generic
  Spacetimes},'' \href{http://xxx.lanl.gov/abs/2110.15356}{{\tt 2110.15356}}.

\bibitem{ArmstrongLM20}
C.~Armstrong, A.~E. Lipstein, and J.~Mei, ``{Color/kinematics duality in
  AdS$_{4}$},'' {\em JHEP} {\bf 02} (2021) 194,
  \href{http://xxx.lanl.gov/abs/2012.02059}{{\tt 2012.02059}}.

\bibitem{Alday:2022lkk}
L.~F. Alday, V.~Gon\c{c}alves, and X.~Zhou, ``{Super Gluon Five-Point
  Amplitudes in AdS Space},'' \href{http://xxx.lanl.gov/abs/2201.04422}{{\tt
  2201.04422}}.

\bibitem{HRTpreprint}
A.~Herderschee, R.~Roiban, and F.~Teng, ``{On the Differential Representation
  and Color-Kinematics Duality of AdS Boundary Correlators},'' {\em to appear
  concurrently} (2022).

\bibitem{Boulware:1981ns}
D.~G. Boulware and L.~S. Brown, ``{Symmetric Space Scalar Field Theory},'' {\em
  Annals Phys.} {\bf 138} (1982) 392.

\bibitem{Camporesi:1990wm}
R.~Camporesi, ``{Harmonic analysis and propagators on homogeneous spaces},''
  {\em Phys. Rept.} {\bf 196} (1990) 1--134.

\bibitem{Boulware:1968zz}
D.~G. Boulware and L.~S. Brown, ``{Tree Graphs and Classical Fields},'' {\em
  Phys. Rev.} {\bf 172} (1968) 1628--1631.

\bibitem{ramond1997field}
P.~Ramond, {\em Field Theory: A Modern Primer}.
\newblock Frontiers in Physics. Avalon Publishing, 1997.

\bibitem{Cheung:2020djz}
C.~Cheung and J.~Mangan, ``{Scattering Amplitudes and the Navier-Stokes
  Equation},'' \href{http://xxx.lanl.gov/abs/2010.15970}{{\tt 2010.15970}}.

\bibitem{Cheung:2020tqz}
C.~Cheung and Z.~Moss, ``{Symmetry and Unification from Soft Theorems and
  Unitarity},'' {\em JHEP} {\bf 05} (2021) 161,
  \href{http://xxx.lanl.gov/abs/2012.13076}{{\tt 2012.13076}}.

\bibitem{Cheung:2021yog}
C.~Cheung, A.~Helset, and J.~Parra-Martinez, ``{Geometric Soft Theorems},''
  \href{http://xxx.lanl.gov/abs/2111.03045}{{\tt 2111.03045}}.

\bibitem{Cheung:2014dqa}
C.~Cheung, K.~Kampf, J.~Novotny, and J.~Trnka, ``{Effective Field Theories from
  Soft Limits of Scattering Amplitudes},'' {\em Phys. Rev. Lett.} {\bf 114}
  (2015), no.~22 221602, \href{http://xxx.lanl.gov/abs/1412.4095}{{\tt
  1412.4095}}.

\bibitem{Hinterbichler:2015pqa}
K.~Hinterbichler and A.~Joyce, ``{Hidden symmetry of the Galileon},'' {\em
  Phys. Rev. D} {\bf 92} (2015), no.~2 023503,
  \href{http://xxx.lanl.gov/abs/1501.07600}{{\tt 1501.07600}}.

\bibitem{BonifacioHJR18}
J.~Bonifacio, K.~Hinterbichler, A.~Joyce, and R.~A. Rosen, ``{Shift Symmetries
  in (Anti) de Sitter Space},'' {\em JHEP} {\bf 02} (2019) 178,
  \href{http://xxx.lanl.gov/abs/1812.08167}{{\tt 1812.08167}}.

\bibitem{BonifacioHJR21}
J.~Bonifacio, K.~Hinterbichler, A.~Joyce, and D.~Roest, ``{Exceptional scalar
  theories in de Sitter space},''
  \href{http://xxx.lanl.gov/abs/2112.12151}{{\tt 2112.12151}}.

\bibitem{Brown:2016hck}
R.~W. Brown and S.~G. Naculich, ``{Color-factor symmetry and BCJ relations for
  QCD amplitudes},'' {\em JHEP} {\bf 11} (2016) 060,
  \href{http://xxx.lanl.gov/abs/1608.05291}{{\tt 1608.05291}}.

\bibitem{DDM}
V.~Del~Duca, L.~J. Dixon, and F.~Maltoni, ``{New color decompositions for gauge
  amplitudes at tree and loop level},'' {\em Nucl. Phys. B} {\bf 571} (2000)
  51--70, \href{http://xxx.lanl.gov/abs/hep-ph/9910563}{{\tt hep-ph/9910563}}.

\bibitem{Herderschee21}
A.~Herderschee, ``{A New Framework for Higher Loop Witten Diagrams},''
  \href{http://xxx.lanl.gov/abs/2112.08226}{{\tt 2112.08226}}.

\bibitem{Gomez:2021ujt}
H.~Gomez, R.~L. Jusinskas, and A.~Lipstein, ``{Cosmological Scattering
  Equations at Tree-level and One-loop},''
  \href{http://xxx.lanl.gov/abs/2112.12695}{{\tt 2112.12695}}.

\bibitem{Paulos11}
M.~F. Paulos, ``{Towards Feynman rules for Mellin amplitudes},'' {\em JHEP}
  {\bf 10} (2011) 074, \href{http://xxx.lanl.gov/abs/1107.1504}{{\tt
  1107.1504}}.

\bibitem{Parikh19}
S.~Parikh, ``{Holographic dual of the five-point conformal block},'' {\em JHEP}
  {\bf 05} (2019) 051, \href{http://xxx.lanl.gov/abs/1901.01267}{{\tt
  1901.01267}}.

\bibitem{JepsenP19}
C.~B. Jepsen and S.~Parikh, ``{Propagator identities, holographic conformal
  blocks, and higher-point AdS diagrams},'' {\em JHEP} {\bf 10} (2019) 268,
  \href{http://xxx.lanl.gov/abs/1906.08405}{{\tt 1906.08405}}.

\bibitem{AlbayrakCK20}
S.~Albayrak, C.~Chowdhury, and S.~Kharel, ``{Study of momentum space scalar
  amplitudes in AdS spacetime},'' {\em Phys. Rev. D} {\bf 101} (2020), no.~12
  124043, \href{http://xxx.lanl.gov/abs/2001.06777}{{\tt 2001.06777}}.

\bibitem{Parikh19B}
S.~Parikh, ``{A multipoint conformal block chain in $d$ dimensions},'' {\em
  JHEP} {\bf 05} (2020) 120, \href{http://xxx.lanl.gov/abs/1911.09190}{{\tt
  1911.09190}}.

\bibitem{KarateevKS17}
D.~Karateev, P.~Kravchuk, and D.~Simmons-Duffin, ``{Weight Shifting Operators
  and Conformal Blocks},'' {\em JHEP} {\bf 02} (2018) 081,
  \href{http://xxx.lanl.gov/abs/1706.07813}{{\tt 1706.07813}}.

\bibitem{CostaH18}
M.~S. Costa and T.~Hansen, ``{AdS Weight Shifting Operators},'' {\em JHEP} {\bf
  09} (2018) 040, \href{http://xxx.lanl.gov/abs/1805.01492}{{\tt 1805.01492}}.

\bibitem{Arkani-HamedBLP18}
N.~Arkani-Hamed, D.~Baumann, H.~Lee, and G.~L. Pimentel, ``{The Cosmological
  Bootstrap: Inflationary Correlators from Symmetries and Singularities},''
  {\em JHEP} {\bf 04} (2020) 105,
  \href{http://xxx.lanl.gov/abs/1811.00024}{{\tt 1811.00024}}.

\bibitem{BaumannDJLP19}
D.~Baumann, C.~Duaso~Pueyo, A.~Joyce, H.~Lee, and G.~L. Pimentel, ``{The
  cosmological bootstrap: weight-shifting operators and scalar seeds},'' {\em
  JHEP} {\bf 12} (2020) 204, \href{http://xxx.lanl.gov/abs/1910.14051}{{\tt
  1910.14051}}.

\bibitem{BaumannDJLP20}
D.~Baumann, C.~Duaso~Pueyo, A.~Joyce, H.~Lee, and G.~L. Pimentel, ``{The
  Cosmological Bootstrap: Spinning Correlators from Symmetries and
  Factorization},'' {\em SciPost Phys.} {\bf 11} (2021) 071,
  \href{http://xxx.lanl.gov/abs/2005.04234}{{\tt 2005.04234}}.

\bibitem{HillmanP21}
A.~Hillman and E.~Pajer, ``{A Differential Representation of Cosmological
  Wavefunctions},'' \href{http://xxx.lanl.gov/abs/2112.01619}{{\tt
  2112.01619}}.

\bibitem{Roehrig:2020kck}
K.~Roehrig and D.~Skinner, ``{Ambitwistor Strings and the Scattering Equations
  on AdS$_3\times$S$^3$},'' \href{http://xxx.lanl.gov/abs/2007.07234}{{\tt
  2007.07234}}.

\bibitem{Penedones:2016voo}
J.~Penedones, ``{TASI lectures on AdS/CFT},'' in {\em {Theoretical Advanced
  Study Institute in Elementary Particle Physics}: {New Frontiers in Fields and
  Strings}}, pp.~75--136, 2017.
\newblock \href{http://xxx.lanl.gov/abs/1608.04948}{{\tt 1608.04948}}.

\end{thebibliography}\endgroup

\end{document}